\documentclass[twocolumn,letterpaper,aps,prc,longbibliography,superscriptaddress,nofootinbib,floatfix]{revtex4-2}

\usepackage{graphicx}	
\usepackage{xspace}     
\usepackage{amsmath}     
\usepackage{multirow}

\def\aleq{\ensuremath{\,\hbox{\lower0.4ex\hbox{$\approx$}\llap{\raise0.5ex\hbox{$<$}}}\,}}
\def\ageq{\ensuremath{\,\hbox{\lower0.4ex\hbox{$\approx$}\llap{\raise0.5ex\hbox{$>$}}}\,}}
\newcommand{\mT}{m_T}
\newcommand{\pT}{p_T}
\newcommand{\uT}{u_T}
\newcommand{\etaprime}{\eta'}
\newcommand{\Tf}{T_f}
\newcommand{\Teff}{T_{\rm eff}}
\newcommand{\Npart}{N_{\rm part}}
\newcommand{\avgalpha}{\alpha_0}
\newcommand{\pp}{$p$$+$$p$}
\newcommand{\ppb}{$p$$+$Pb }
\newcommand{\pbpb}{Pb$+$Pb }
\newcommand{\sqsntwo}{$\sqrt{s_{_{NN}}}$}

\newcommand{\epem}{{e$^+$}\kern-.15em{+e$^-$}}

\begin{document}

\title{Centrality dependence of L\'evy-stable two-pion Bose-Einstein
correlations in $\sqrt{s_{_{NN}}}=200$ GeV Au$+$Au collisions.}

\newcommand{\abilene}{Abilene Christian University, Abilene, Texas 79699, USA}
\newcommand{\augie}{Department of Physics, Augustana University, Sioux Falls, South Dakota 57197, USA}
\newcommand{\banaras}{Department of Physics, Banaras Hindu University, Varanasi 221005, India}
\newcommand{\barc}{Bhabha Atomic Research Centre, Bombay 400 085, India}
\newcommand{\baruch}{Baruch College, City University of New York, New York, New York, 10010 USA}
\newcommand{\bnlcoll}{Collider-Accelerator Department, Brookhaven National Laboratory, Upton, New York 11973-5000, USA}
\newcommand{\bnlphys}{Physics Department, Brookhaven National Laboratory, Upton, New York 11973-5000, USA}
\newcommand{\caucr}{University of California-Riverside, Riverside, California 92521, USA}
\newcommand{\charlesczech}{Charles University, Faculty of Mathematics and Physics, 180 00 Troja, Prague, Czech Republic}
\newcommand{\cns}{Center for Nuclear Study, Graduate School of Science, University of Tokyo, 7-3-1 Hongo, Bunkyo, Tokyo 113-0033, Japan}
\newcommand{\colorado}{University of Colorado, Boulder, Colorado 80309, USA}
\newcommand{\columbia}{Columbia University, New York, New York 10027 and Nevis Laboratories, Irvington, New York 10533, USA}
\newcommand{\czechtech}{Czech Technical University, Zikova 4, 166 36 Prague 6, Czech Republic}
\newcommand{\dapnia}{Dapnia, CEA Saclay, F-91191, Gif-sur-Yvette, France}
\newcommand{\debrecen}{Debrecen University, H-4010 Debrecen, Egyetem t{\'e}r 1, Hungary}
\newcommand{\elte}{ELTE, E{\"o}tv{\"o}s Lor{\'a}nd University, H-1117 Budapest, P{\'a}zm{\'a}ny P.~s.~1/A, Hungary}
\newcommand{\ewha}{Ewha Womans University, Seoul 120-750, Korea}
\newcommand{\fsu}{Florida State University, Tallahassee, Florida 32306, USA}
\newcommand{\gsu}{Georgia State University, Atlanta, Georgia 30303, USA}
\newcommand{\hanyang}{Hanyang University, Seoul 133-792, Korea}
\newcommand{\hiroshima}{Physics Program and International Institute for Sustainability with Knotted Chiral Meta Matter (WPI-SKCM$^2$), Hiroshima University, Higashi-Hiroshima, Hiroshima 739-8526, Japan}
\newcommand{\hunrenatomki}{HUN-REN ATOMKI, H-4026 Debrecen, Bem t{\'e}r 18/c, Hungary}
\newcommand{\ihepprot}{IHEP Protvino, State Research Center of Russian Federation, Institute for High Energy Physics, Protvino, 142281, Russia}
\newcommand{\illuiuc}{University of Illinois at Urbana-Champaign, Urbana, Illinois 61801, USA}
\newcommand{\inrras}{Institute for Nuclear Research of the Russian Academy of Sciences, prospekt 60-letiya Oktyabrya 7a, Moscow 117312, Russia}
\newcommand{\instpasczech}{Institute of Physics, Academy of Sciences of the Czech Republic, Na Slovance 2, 182 21 Prague 8, Czech Republic}
\newcommand{\isu}{Iowa State University, Ames, Iowa 50011, USA}
\newcommand{\jaea}{Advanced Science Research Center, Japan Atomic Energy Agency, 2-4 Shirakata Shirane, Tokai-mura, Naka-gun, Ibaraki-ken 319-1195, Japan}
\newcommand{\jeonbuk}{Jeonbuk National University, Jeonju, 54896, Korea}
\newcommand{\jyvaskyla}{Helsinki Institute of Physics and University of Jyv{\"a}skyl{\"a}, P.O.Box 35, FI-40014 Jyv{\"a}skyl{\"a}, Finland}
\newcommand{\kek}{KEK, High Energy Accelerator Research Organization, Tsukuba, Ibaraki 305-0801, Japan}
\newcommand{\korea}{Korea University, Seoul 02841, Korea}
\newcommand{\kurchatov}{National Research Center ``Kurchatov Institute", Moscow, 123098 Russia}
\newcommand{\kyoto}{Kyoto University, Kyoto 606-8502, Japan}
\newcommand{\labllr}{Laboratoire Leprince-Ringuet, Ecole Polytechnique, CNRS-IN2P3, Route de Saclay, F-91128, Palaiseau, France}
\newcommand{\lahorelums}{Physics Department, Lahore University of Management Sciences, Lahore 54792, Pakistan}
\newcommand{\lawllnl}{Lawrence Livermore National Laboratory, Livermore, California 94550, USA}
\newcommand{\losalamos}{Los Alamos National Laboratory, Los Alamos, New Mexico 87545, USA}
\newcommand{\lpc}{LPC, Universit{\'e} Blaise Pascal, CNRS-IN2P3, Clermont-Fd, 63177 Aubiere Cedex, France}
\newcommand{\lund}{Department of Physics, Lund University, Box 118, SE-221 00 Lund, Sweden}
\newcommand{\maryland}{University of Maryland, College Park, Maryland 20742, USA}
\newcommand{\mass}{Department of Physics, University of Massachusetts, Amherst, Massachusetts 01003-9337, USA}
\newcommand{\mate}{MATE, Institute of Technology, Laboratory of Femtoscopy, K\'aroly R\'obert Campus, H-3200 Gy\"ongy\"os, M\'atrai \'ut 36, Hungary}
\newcommand{\michigan}{Department of Physics, University of Michigan, Ann Arbor, Michigan 48109-1040, USA}
\newcommand{\miss}{Mississippi State University, Mississippi State, Mississippi 39762, USA}
\newcommand{\muhlenberg}{Muhlenberg College, Allentown, Pennsylvania 18104-5586, USA}
\newcommand{\myongji}{Myongji University, Yongin, Kyonggido 449-728, Korea}
\newcommand{\nagasaki}{Nagasaki Institute of Applied Science, Nagasaki-shi, Nagasaki 851-0193, Japan}
\newcommand{\nara}{Nara Women's University, Kita-uoya Nishi-machi Nara 630-8506, Japan}
\newcommand{\natmephi}{National Research Nuclear University, MEPhI, Moscow Engineering Physics Institute, Moscow, 115409, Russia}
\newcommand{\newmex}{University of New Mexico, Albuquerque, New Mexico 87131, USA}
\newcommand{\nmsu}{New Mexico State University, Las Cruces, New Mexico 88003, USA}
\newcommand{\northcg}{Physics and Astronomy Department, University of North Carolina at Greensboro, Greensboro, North Carolina 27412, USA}
\newcommand{\ohio}{Department of Physics and Astronomy, Ohio University, Athens, Ohio 45701, USA}
\newcommand{\ornl}{Oak Ridge National Laboratory, Oak Ridge, Tennessee 37831, USA}
\newcommand{\orsay}{IPN-Orsay, Univ.~Paris-Sud, CNRS/IN2P3, Universit\'e Paris-Saclay, BP1, F-91406, Orsay, France}
\newcommand{\pnpi}{PNPI, Petersburg Nuclear Physics Institute, Gatchina, Leningrad region, 188300, Russia}
\newcommand{\riken}{RIKEN Nishina Center for Accelerator-Based Science, Wako, Saitama 351-0198, Japan}
\newcommand{\rikjrbrc}{RIKEN BNL Research Center, Brookhaven National Laboratory, Upton, New York 11973-5000, USA}
\newcommand{\rikkyo}{Physics Department, Rikkyo University, 3-34-1 Nishi-Ikebukuro, Toshima, Tokyo 171-8501, Japan}
\newcommand{\saispbstu}{Saint Petersburg State Polytechnic University, St.~Petersburg, 195251 Russia}
\newcommand{\saopaulo}{Universidade de S{\~a}o Paulo, Instituto de F\'{\i}sica, Caixa Postal 66318, S{\~a}o Paulo CEP05315-970, Brazil}
\newcommand{\seoulnat}{Department of Physics and Astronomy, Seoul National University, Seoul 151-742, Korea}
\newcommand{\stonybrkc}{Chemistry Department, Stony Brook University, SUNY, Stony Brook, New York 11794-3400, USA}
\newcommand{\stonycrkp}{Department of Physics and Astronomy, Stony Brook University, SUNY, Stony Brook, New York 11794-3800, USA}
\newcommand{\sungskku}{Sungkyunkwan University, Suwon, 440-746, Korea}
\newcommand{\tenn}{University of Tennessee, Knoxville, Tennessee 37996, USA}
\newcommand{\titech}{Department of Physics, Tokyo Institute of Technology, Oh-okayama, Meguro, Tokyo 152-8551, Japan}
\newcommand{\tsukuba}{Tomonaga Center for the History of the Universe, University of Tsukuba, Tsukuba, Ibaraki 305, Japan}
\newcommand{\usmma}{United States Merchant Marine Academy, Kings Point, New York 11024, USA}
\newcommand{\vandy}{Vanderbilt University, Nashville, Tennessee 37235, USA}
\newcommand{\weizmann}{Weizmann Institute, Rehovot 76100, Israel}
\newcommand{\wigner}{Institute for Particle and Nuclear Physics, HUN-REN Wigner Research Centre for Physics, (HUN-REN Wigner RCP, RMI), H-1525 Budapest 114, POBox 49, Budapest, Hungary}
\newcommand{\yonsei}{Yonsei University, IPAP, Seoul 120-749, Korea}
\newcommand{\zagreb}{Department of Physics, Faculty of Science, University of Zagreb, Bijeni\v{c}ka c.~32 HR-10002 Zagreb, Croatia}
\newcommand{\zambia}{Department of Physics, School of Natural Sciences, University of Zambia, Great East Road Campus, Box 32379, Lusaka, Zambia}
\affiliation{\abilene}
\affiliation{\augie}
\affiliation{\banaras}
\affiliation{\barc}
\affiliation{\baruch}
\affiliation{\bnlcoll}
\affiliation{\bnlphys}
\affiliation{\caucr}
\affiliation{\charlesczech}
\affiliation{\cns}
\affiliation{\colorado}
\affiliation{\columbia}
\affiliation{\czechtech}
\affiliation{\dapnia}
\affiliation{\debrecen}
\affiliation{\elte}
\affiliation{\ewha}
\affiliation{\fsu}
\affiliation{\gsu}
\affiliation{\hanyang}
\affiliation{\hiroshima}
\affiliation{\hunrenatomki}
\affiliation{\ihepprot}
\affiliation{\illuiuc}
\affiliation{\inrras}
\affiliation{\instpasczech}
\affiliation{\isu}
\affiliation{\jaea}
\affiliation{\jeonbuk}
\affiliation{\jyvaskyla}
\affiliation{\kek}
\affiliation{\korea}
\affiliation{\kurchatov}
\affiliation{\kyoto}
\affiliation{\labllr}
\affiliation{\lahorelums}
\affiliation{\lawllnl}
\affiliation{\losalamos}
\affiliation{\lpc}
\affiliation{\lund}
\affiliation{\maryland}
\affiliation{\mass}
\affiliation{\mate}
\affiliation{\michigan}
\affiliation{\miss}
\affiliation{\muhlenberg}
\affiliation{\myongji}
\affiliation{\nagasaki}
\affiliation{\nara}
\affiliation{\natmephi}
\affiliation{\newmex}
\affiliation{\nmsu}
\affiliation{\northcg}
\affiliation{\ohio}
\affiliation{\ornl}
\affiliation{\orsay}
\affiliation{\pnpi}
\affiliation{\riken}
\affiliation{\rikjrbrc}
\affiliation{\rikkyo}
\affiliation{\saispbstu}
\affiliation{\saopaulo}
\affiliation{\seoulnat}
\affiliation{\stonybrkc}
\affiliation{\stonycrkp}
\affiliation{\sungskku}
\affiliation{\tenn}
\affiliation{\titech}
\affiliation{\tsukuba}
\affiliation{\usmma}
\affiliation{\vandy}
\affiliation{\weizmann}
\affiliation{\wigner}
\affiliation{\yonsei}
\affiliation{\zagreb}
\affiliation{\zambia}
\author{N.J.~Abdulameer} \affiliation{\debrecen} \affiliation{\hunrenatomki}
\author{U.~Acharya} \affiliation{\gsu}
\author{A.~Adare} \affiliation{\colorado} 
\author{C.~Aidala} \affiliation{\losalamos} \affiliation{\michigan} 
\author{N.N.~Ajitanand} \altaffiliation{Deceased} \affiliation{\stonybrkc} 
\author{Y.~Akiba} \email[PHENIX Spokesperson: ]{akiba@rcf.rhic.bnl.gov} \affiliation{\riken} \affiliation{\rikjrbrc}
\author{R.~Akimoto} \affiliation{\cns} 
\author{H.~Al-Ta'ani} \affiliation{\nmsu} 
\author{J.~Alexander} \affiliation{\stonybrkc} 
\author{A.~Angerami} \affiliation{\columbia} 
% ...
\collaboration{PHENIX Collaboration}  \noaffiliation

\date{\today}


\begin{abstract}


The PHENIX experiment measured the centrality dependence of two-pion
Bose-Einstein correlation functions in $\sqrt{s_{_{NN}}}=200$~GeV Au$+$Au
collisions at the Relativistic Heavy Ion Collider at Brookhaven National
Laboratory. The data are well represented by L\'evy-stable source
distributions. The extracted source parameters are the
correlation-strength parameter $\lambda$, the L\'evy index of stability
$\alpha$, and the L\'evy-scale parameter $R$ as a function of transverse
mass $m_T$ and centrality. The $\lambda(m_T)$ parameter is constant at
larger values of $m_T$, but decreases as $m_T$ decreases. The L\'evy scale
parameter $R(m_T)$ decreases with $m_T$ and exhibits proportionality to
the length scale of the nuclear overlap region. The L\'evy exponent
$\alpha(m_T)$ is independent of $m_T$ within uncertainties in each
investigated centrality bin, but shows a clear centrality dependence. At
all centralities, the L\'evy exponent $\alpha$ is significantly different
from that of Gaussian ($\alpha=2$) or Cauchy ($\alpha=1$) source
distributions. Comparisons to the predictions of Monte-Carlo simulations
of resonance-decay chains show that in all but the most-peripheral
centrality class (50\%--60\%), the obtained results are inconsistent with
the measurements, unless a significant reduction of the in-medium mass of
the $\eta'$ meson is included. In each centrality class, the best value of
the in-medium $\eta'$ mass is compared to the mass of the $\eta$ meson, as
well as to several theoretical predictions that consider restoration of
$U_A(1)$ symmetry in hot hadronic matter.

\end{abstract}

\maketitle

\section{Introduction}

In a previous paper~\cite{PHENIX:2017ino} on Bose-Einstein Correlations 
(BECs)---also known as Hanbury Brown and Twiss (HBT) Correlations---the 
PHENIX Collaboration found that for 0\%--30\% centrality Au$+$Au 
collisions at \sqsntwo=200 GeV, the two-particle BECs are well-described 
by a L\'evy-stable source distribution.  However, the traditional 
description of the same dataset, using a Gaussian source distribution 
was found to be inadequate~\cite{PHENIX:2017ino}. A strong preference 
for the L\'evy description had also been seen in \epem collisions at the 
Large Electron-Positron Collider~\cite{L3:2011kzb} and in \pp, \ppb, and 
\pbpb collisions at the Large Hadron 
Collider~\cite{CMS:2023xyd,CMS:2011nlc,Sikler:2014aea,ATLAS:2015dqi}, in 
Be+Be and Ar+Sc collisions at the Super Proton 
Synchrotron~\cite{NA61SHINE:2023qzr,Porfy:2024kbk} and in Au$+$Au 
collisions at the Relativistic Heavy Ion Collider 
(RHIC)~\cite{Kincses:2024sin}.

Presented here is a precise measurement of the centrality and 
transverse-mass dependence of the two-pion BEC function in Au$+$Au 
collisions at \sqsntwo = 200 GeV by the PHENIX experiment at RHIC. This 
data sample, recorded in 2010, allows a fine transverse-mass binning and 
inference of the shape of the correlation function more precisely than 
was possible with earlier data sets. For the first time, these results 
are presented as a function of centrality for six centrality 
classes in the range 0\%--60\%. As was done for the 0\%--30\% centrality 
class in Ref.~\cite{PHENIX:2017ino}, the source parameters of the L\'evy 
distribution ($\lambda$, $R$, $\alpha$) are measured.\footnote{Note that 
followed are the conventions introduced in Refs.~\cite{Csorgo:2003uv} 
and ~\cite{Csorgo:2004sr,Novak:2016cyc,Csorgo:2018uyp}, which is based 
on a book by P. J. Nolan~\cite{Nolan:2020abc} on univariate 
L\'evy-stable source distributions, including also multivariate, but 
symmetric L\'evy-stable distributions.}.
The centrality and the transverse-mass dependence of the L\'evy-fit 
parameters are characterized with simple, theoretically and empirically 
motivated fit functions. The centrality dependence of the parameters of 
these functions is investigated in detail.

The structure of this paper is as follows: Sections~\ref{sec:phenix} 
and~\ref{sec:data} present the PHENIX experimental setup and the 
selection of the data sample, respectively. Section~\ref{sec:analysis} 
explores the procedure of measurement and fitting of the two-pion 
correlation function. Section~\ref{sec:systematics} discusses the 
systematic uncertainties. Section~\ref{sec:results} presents the 
extracted L\'evy parameters of the source as a function of  
centrality. Section~\ref{sec:simulations} discusses Monte Carlo 
simulations and some of the possible physics interpretations of these 
results, which have a strong exclusion power due to their high 
precision. Section~\ref{sec:conclusions} summarizes and concludes. 
Finally an Appendix details our Monte-Carlo simulations to interpret the 
PHENIX data.

In particular, PHENIX data on the transverse-mass and centrality 
dependence of the L\'evy intercept parameter $\lambda(\mT)$ are compared 
to centrality-dependent Monte-Carlo simulations of resonance decay 
chains. In all but the most-peripheral centrality class (50\%--60\%), 
the Monte-Carlo simulations are found to be inconsistent with the 
measurements unless a significant reduction of the in-medium mass of the 
$\eta'$ meson is included. In each centrality class, the best value of 
the in-medium $\eta'$ mass is determined from $\chi^2$ and confidence 
level (CL or p-value) maps, based on a comparison of PHENIX data and 
Monte-Carlo simulations. The resulting values of in-medium modified 
$\etaprime$ masses are compared to the mass of the $\eta$ meson as well 
as to several theoretical predictions that consider restoration of 
$U_A(1)$ symmetry in hot hadronic matter as discussed in 
Sections~\ref{sec:simulations} and~\ref{sec:conclusions} and then 
further detailed in the Appendix. 
Throughout this paper, units are used such that $\hbar=c=1$. Also, the 
utilized fits represent the fitted data with CL in the statistically 
allowed 0.1\%$\leq$CL$\leq$99.9\% interval.

\section{The PHENIX experiment}
\label{sec:phenix}

The data used in this analysis are the same, apart from the centrality 
selection, as in the previous PHENIX L\'evy HBT analysis with a 
0\%--30\% centrality selection~\cite{PHENIX:2017ino}. The PHENIX 
experimental apparatus relevant to this analysis is thus also the same. 
Briefly, the PHENIX detector is subdivided into the central-arm 
spectrometer (covering 2$\times$90$^\circ$ azimuthal and $|\eta| \le 
0.35$ pseudorapidity acceptance), which is used here to focus mainly on 
hadron, electron, and photon identification and measurement.  In the 
forward direction for each beam, two muon-arm spectrometers are used to 
focus mainly on identification and measurement of muons. There are also 
various event-characterization and triggering detectors in place. Of 
particular benefit here is good identification and measurement of 
charged pions.  Ref.~\cite{PHENIX:2017ino} provides further details.

\section{Data sample}
\label{sec:data}

The data sample used in this analysis comprises Au$+$Au collisions 
recorded by the PHENIX detector at \sqsntwo = 200 GeV in 2010. The 
minimum-bias data sample contains $\approx$7.3 billion events which is 
reduced to $\approx$4.4 billion with the 0\%--60\% centrality selection. 
The centrality dependence of the transverse-mass trends is explored 
here in term of the numbers of participants ($\Npart$), which was 
determined via Glauber-model calculations by the PHENIX experiment 
based on Ref.~\cite{Miller:2007ri}.

The present analysis shares almost all details with the previous 
analysis of Ref.~\cite{PHENIX:2017ino}, including using the word ``cuts" 
to refer to selection criteria. The similarities and differences between 
the current and previous analyses are detailed below.  In the present 
analysis, the L\'evy fit parameters are determined in 23 bins of 
transverse mass from 0.248 GeV to 0.876 GeV and in six, 10\% wide 
centrality bins in the range of 0\%--60\%. Well-measured tracks are 
selected using the same single-track cuts as in 
Ref.~\cite{PHENIX:2017ino}. The event-selection criteria (except the 
centrality selection) and the particle-identification techniques are 
also the same as in Ref.~\cite{PHENIX:2017ino}. The single-track cuts 
and their variations are considered as sources of systematic 
uncertainties, as described in Section~\ref{sec:systematics}.

The particle identification (PID) of pions is based on time-of-flight 
information and the path length information given by the track model. As 
in Ref.~\cite{PHENIX:2017ino}, a general cut on 
transverse momentum, $p_T>0.16$ GeV, is applied to all pions. The cuts 
used in the PID are also considered as sources of systematic 
uncertainties.


In addition to the cuts on single tracks, pair cuts are imposed to 
minimize two-track effects: track merging, and splitting. Merging occurs 
when two tracks are so close to each other that the reconstruction 
algorithm considers them to be one track. Splitting is the opposite of 
the merging effect: one track is falsely reconstructed as two. These 
ambiguous pairs can be removed from the sample by geometrical cuts on 
their $\Delta \varphi$--$\Delta z$ plane, where $\Delta \varphi$ denotes 
the azimuthal angle difference of the hit positions and $\Delta z$ is 
the difference of the $z$ coordinates of the pair, as determined by the 
drift chambers (DC), lead-scintillator electromagnetic calorimeter 
(PbSc), and time-of-flight (TOF) in the east and west arms of the PHENIX 
spectrometer.

These pair cuts were carefully investigated in the previous L\'evy 
analysis~\cite{PHENIX:2017ino}. However, due to the different centrality 
selections, the pair-cut settings here are slightly modified.  The pair 
cuts are defined in the $\Delta \varphi$--$\Delta z$ plane as:
\begin{align}
&\Delta \varphi > \Delta \varphi_0 \left( 1{-}\frac{\Delta z}{\Delta z_0} \right) \hspace{0.15cm} {\rm and} \hspace{0.15cm} \Delta \varphi > \Delta \varphi_1 \hspace{0.2cm} 
\text{(DC and PbSc)}, \\
&\Delta \varphi > \Delta \varphi_0 \left( 1{-}\frac{\Delta z}{\Delta z_0} \right) \hspace{0.2cm} \text{(TOF east)}, \\
&\Delta \varphi > \Delta \varphi_0 \hspace{0.15cm} {\rm and} \hspace{0.15cm} \Delta z > \Delta z_0 \hspace{0.15cm} \text{(TOF west)}.
\end{align}
The default values of the $\Delta \varphi_0$, $\Delta \varphi_1$, and 
$\Delta z_0$ can be found in Table~\ref{tab:paircuts-alt}, where also 
listed are the alternative values, which are used in 
Section~\ref{sec:systematics} to determine the systematic uncertainties.

\begin{table*}[ht!]
\caption{\label{tab:paircuts-alt} 
The values of the coordinates for the pair-selection (cuts) criteria  
and the alternative values used to determine systematic uncertainties.
}
\begin{ruledtabular} \begin{tabular}{ccccccccccccccccccc}
\null & \multicolumn{5}{c}{DC} & \multicolumn{4}{c}{TOF east} 
& \multicolumn{4}{c}{TOF west}  & \multicolumn{5}{c}{EM Cal} \\ 
\multirow{1}{2em}{Pair cuts} 
&& $\Delta \varphi_0$ & $\Delta z_0$ & $\Delta \varphi_1$ 
&&& $\Delta \varphi_0$ & $\Delta z_0$ 
&&& $\Delta \varphi_0$ & $\Delta z_0$ 
&&& $\Delta \varphi_0$ & $\Delta z_0$ & $\Delta \varphi_1$ & \\
\null                    &&[rad]&[cm]& [rad]&&& [rad]&[cm] &&& [rad] &[cm]  &&&[rad]&[cm]& [rad]&\\
\hline
Default cut settings     && 0.12 & 8. & 0.017 &&&  0.12 & 12  &&&  0.075 & 14.0 &&& 0.12 & 16 & 0.015 & \\
Loose drift chamber cut  && 0.11 & 7. & 0.016 &&&  0.12 & 12  &&&  0.075 & 14.0 &&& 0.12 & 16 & 0.015 & \\
Strict drift chamber cut && 0.13 & 9. & 0.018 &&&  0.12 & 12  &&&  0.075 & 14.0 &&& 0.12 & 16 & 0.015 & \\
Loose ID detector cuts   && 0.12 & 8. & 0.017 &&&  0.11 & 11  &&&  0.070 & 13.0 &&& 0.11 & 15 & 0.013 & \\
Strict ID detector cuts  && 0.12 & 8. & 0.017 &&&  0.13 & 13  &&&  0.080 & 15.0 &&& 0.13 & 17 & 0.017 & \\
\end{tabular} \end{ruledtabular} 
\end{table*}

As in Ref.~\cite{PHENIX:2017ino}, in addition to these cuts, if multiple 
tracks are found that are associated with hits in the same tower of the 
PbSc slat of the TOF east, or strip of the TOF west detector, then all 
but one (randomly chosen) are removed. This ensures that no ghost tracks 
remain in the sample after the above-mentioned pair cuts.

Within statistical uncertainties of the fit parameters, using only 
positive pions gives results consistent with using only negative pions.  
Consequently, only the results from combined fits to both $(++)$ and 
$(--)$ charge combinations of identified pion pairs are presented here.


\section{Measuring and fitting the two-particle correlation function}
\label{sec:analysis}

\subsection{Measuring the correlation function}

In principle, a detailed shape analysis of the two-particle correlation 
functions could require three-dimensional measurements, but the lack of 
statistical precision could make such measurements impractical.
The goal here is to obtain precise results in several transverse-mass 
and centrality bins. Thus, the correlation functions use a single 
variable $Q$.  PHENIX preliminary results on a multivariate L\'evy 
analysis are also available~\cite{Kurgyis:2018zck}, but go beyond the 
scope of the present paper.

Our relative momentum variable $Q$ is chosen as the modulus of the 
three-momentum difference in the longitudinal comoving system 
(LCMS)~\cite{Pratt:1990zq}. With the Bertsch-Pratt decomposition of the 
relative momentum to the ``side, out, and long" 
components~\cite{Bertsch:1988db,Pratt:1986cc} in the LCMS, the variable 
can be written as
\begin{align}
    Q &\equiv |\textbf{q}_{\rm LCMS}| = \sqrt{q^2_{\rm out,LCMS} + q^2_{\rm side,LCMS}+q^2_{\rm long,LCMS}}.
\end{align}
The motivation for this variable comes from the experimental 
observation~\cite{Csorgo:2003uv} that in a Gaussian three-dimensional 
BEC analysis, a quadratic sum appearing in the 
BEC functions provided nearly equal BEC radius 
parameters, i.e, $R_i {\approx}R$ for $i = {\rm side, out, long}$.
Thus, this quadratic sum can be simplified as
\begin{align}
    \sum_{i = {\rm side, out, long}} R_i^2 q_i^2 {\approx}R^2 \left( \sum_{i = {\rm side, out, long}} q_i^2 \right) = R^2 Q^2,
\end{align}

\noindent which depends on the relative momentum only through the 
one-dimensional variable $Q$, the magnitude of the relative momentum of 
the pair in the LCMS. It was shown in Ref.~\cite{Csorgo:2003uv} that the 
same quadratic sum appears in three-dimensional symmetric 
L{\'e}vy-stable distributions. So our choice for the relative momentum 
variable $Q$ actually includes, as a special case, the L\'evy analysis 
of three-dimensional Bose-Einstein correlation functions under the 
condition that the HBT radii in all three spatial dimensions are equal 
within experimental uncertainties. An approximate equality of the 
transverse-mass and centrality-dependent BEC radii was found in several 
experiments, e.g., S$+$Pb collisions at the CERN 
Super-Proton-Synchrotron energies by the NA44 
Collaboration~\cite{NA44:1994dmh}, and in $\sqrt{s_{NN}} = 200$ GeV 
Au$+$Au collisions at $\sqrt{s_{NN}}=200$~GeV by the PHENIX and STAR 
Collaborations in the Gaussian approximation in 
Refs.~\cite{PHENIX:2004yan,STAR:2004qya}. PHENIX preliminary 
multivariate L\'evy fits in Au$+$Au collisions at 
$\sqrt{s_{NN}}=200$~GeV are also used in Ref.~\cite{Kurgyis:2018zck}. 
Especially note that the PHENIX preliminary three-dimensional 
L\'evy-analysis for Au$+$Au collisions at \sqsntwo=200~GeV finds that 
the three BEC radii in the LCMS are approximately equal (except perhaps 
at small transverse mass).

Furthermore, the radii are approximately equal to the radius found in a 
one-dimensional analysis~\cite{Kurgyis:2018zck}, i.e., the L\'evy source 
is indeed approximately spherical in the LCMS. Hence, this 
one-dimensional L\'evy analysis, which focuses on the centrality and 
transverse-mass dependence, can be considered also as a reasonable 
approximation of a three-dimensional L\'evy analysis. A detailed 
justification of this choice of $Q$ can also be found in 
Ref.~\cite{PHENIX:2017ino}, where the same variable is used.

Our choice of $Q$ can be written using the measured momenta of 
identified pions, $p_i^\mu = (E_i, p_{i,x}, p_{i,y}, p_{i,z})$.  
For $i=1,2$:
\begin{widetext}
    \begin{align}
    Q &\equiv |\textbf{q}_{\rm LCMS}| = \sqrt{(p_{\rm 1,x}-p_{
m 2,x})^2 + (p_{\rm 1,y}-p_{\rm 2,y})^2+\frac{4(p_{\rm 1,z}E_2-p_{\rm 2,z}E_1)^2}{(E_1+E_2)^2-(p_{1,z}+p_{2,z})^2}},
    \label{eq:Qdef}
    \end{align}
\end{widetext}
where the z axis coincides with the beam axis.

The correlation function is measured as:
\begin{align}
C_2(Q) = \frac{A(Q)}{B(Q)}\cdot \frac{\int_{Q_{\rm int,min}}^{Q_{\rm int,max}} B(m_T,Q)}{\int_{Q_{\rm int,min}}^{Q_{\rm int,max}} A(m_T,Q)},    
\label{eq:C2exp}
\end{align}
where $A(Q)$ is the actual $Q$ distribution of pairs of identical pions 
coming from the same event, $B(Q)$ is the corresponding distribution of 
pairs of identical pions from different events, and $Q_{\rm int,min}$ 
and $Q_{\rm int,max}$ denote the lower and upper bound of the integrals. 
Note that $A(Q)$ may contain correlations in addition to the 
quantum-statistical correlation due to the indistinguishability of the 
the measurement of identical bosons.  Other correlations could be the 
consequence of conservation laws, resonance decays, kinematics, detector 
acceptance effects, etc.

Ideally, the background $B(Q)$ distribution, or reference sample, is 
identical to the sample of like-charged pion pairs in all respects, 
except for the Bose-Einstein interference effect 
itself~\cite{Kittel:2005fu}. As such a $B(Q)$ distribution does not 
exist in Nature; it has to be generated using approximation schemes. The 
choice here is explained below.  Careful testing showed stability over 
the choice of the fit range and systematic variations of acceptance, 
PID, and other cuts detailed among the systematic uncertainties.

In choosing the background distribution $B(Q)$, each member of the pairs 
are selected from different events.  Hence, $B(Q)$ contains only trivial 
kinematic correlations between independent particles that are 
distributed with the same single-particle spectra and in the same 
kinematic range. The background distribution is affected by the 
centrality-selection cuts and detector-acceptance effects, rapidity, and 
transverse-mass cuts. In $A(Q)$ the pairs are correlated not only due to 
Bose-Einstein correlations, but also by other initial- or final-state 
interactions. Such effects from interactions include branching processes 
of jets, hadronization effects, resonance decays, energy, and momentum 
conservation laws, possible other kinematic effects, such as elastic 
scattering, Coulomb, and strong final-state interactions.

In general, $A(Q|MC)$ and $B(Q|MC)$ should be generated from Monte-Carlo 
simulations that describe these single-particle data and all other 
nonBEC correlation effects. These Monte-Carlo distributions are then 
used to correct $B(Q)$ for its lack of nonBEC correlations, as was done, 
for example in \epem\ collisions by the L3 
collaboration~\cite{L3:2011kzb}. However, in high-energy heavy-ion 
physics there is a good reason why this complicated and 
Monte-Carlo-dependent procedure is not necessary.  Namely, the presence 
of high-multiplicity events, with $\langle n \rangle \gg 1$, where 
$\langle n \rangle$ is the mean charged-particle multiplicity at 
midrapidity. The usual kinematic correlations, which are due to 
resonance decays or conservation laws, are proportional to the mean 
multiplicity, $\langle n \rangle$, while Bose-Einstein correlations grow 
with the mean number of pairs $\langle n(n-1)\rangle$. Thus in 
high-multiplicity heavy ion collisions, Bose-Einstein correlations 
(together with Coulomb and strong final-state interactions that also 
grow proportionally with the number of pairs) outnumber all the other 
correlations. Hence, the MC approach can be safely abandoned. In the 
expression $A(Q)/B(Q)$ the Bose-Einstein, Coulomb, and strong 
final-state correlations dominate, and the other correlations are 
suppressed as $\langle n\rangle/\langle n(n-1)\rangle \propto 1/\langle 
n \rangle \ll 1$.

In BEC measurements using charged particles, Coulomb repulsion modifies 
the correlation function at low $Q$ values creating the ``Coulomb 
hole"~\cite{Pratt:1990zq}. To account for the final-state Coulomb 
interaction, the Coulomb wave function is integrated over the source of 
pions. Strong final-state interactions could also be taken into account 
using phase shifts that modify the Coulomb wave function. However, the 
strong final-state interaction of the pion pairs is small compared to 
the experimental precision~\cite{Pratt:1990zq,Kincses:2019rug}. All the 
L\'evy fits here (without corrections for the strong final-state 
interactions) are of good quality with CL $\gg 0.1$\%. The strong 
final-state interactions are found not to affect the quality of L\'evy 
fits, nor to change the parameters 
significantly~\cite{Pratt:1990zq,Kincses:2019rug}. Therefore, such 
corrections are not considered in this analysis. In contrast, the effect 
of Coulomb interactions is clearly visible as a Coulomb hole at small 
values of $Q$. The fitting function is appropriately modified, as 
detailed Section~\ref{sec:CCcoul}.

\subsection{The fitting function}
\label{sec:CC}

\subsubsection{The L\'evy shape \label{sec:CCLshape} }

For the shape of the correlation function, it is not possible to 
$a$~$priori$ assume or know, what is or what should be the shape-model 
of the Bose-Einstein correlation functions. Measurements of this 
quantity test the hypothesis of being consistent with a L{\'e}vy shape. 
Using the plane-wave approximation and assuming a spherically symmetric, 
L\'evy-type source, the two-particle BEC function has the simple form
\begin{align}
    C_2(Q) = 1 + \lambda \exp{\left[-Q^\alpha R^\alpha\right]},
    \label{eq:BEC-stable}
\end{align}
where $\lambda$ is the strength of the correlation, $R$ is the scale 
parameter in physical units, and $\alpha$ is the L\'evy index of 
stability~\cite{Csorgo:2003uv}. This hypothesis is tested not only by 
successful fits to the PHENIX data with CL $\gg 0.1$\%, which are stable 
and robust with systematic variations of particle identification, 
acceptance, and pair cuts, as detailed in Section~\ref{sec:systematics}. 
The validity of the L\'evy shape is also checked by employing a L\'evy 
expansion technique~\cite{Csorgo:2000pf,Novak:2016cyc,Csorgo:2018uyp}. 
This method utilizes a complete set of polynomials that are orthonormal 
with respect to a L\'evy weight function~\cite{Novak:2016cyc}; so this 
method is able to characterize and model independently any deviation from a 
L\'evy-stable source shape. To first order of the expansion, no 
significant deviation from the L\'evy shape is found in any of the 
centrality and transverse-mass ranges investigated here.

The relevance of stable distributions to the analysis of 
(Coulomb-corrected) Bose-Einstein correlations was studied in 
Ref.~\cite{Csorgo:2003uv}, following the general mathematical ideas 
summarized in Ref.~\cite{Nolan:2020abc}. Univariate stable distributions 
are usually characterized by the Fourier transform of their density 
distributions that are called the characteristic functions. Following 
the convention of the previous analysis~\cite{PHENIX:2017ino} and the 
theoretical paper where the idea first appeared~\cite{Csorgo:2003uv}, 
the $S(\alpha,\beta,\gamma,\delta;1)$ notation is used. The parameter 
$\alpha$ is the L\'evy index of stability (or characteristic exponent) 
that is limited to the domain $0 < \alpha \leq 2$. The asymmetry 
parameter $\beta$ is limited to the domain $-1 \leq \beta \leq 1$. The 
scale parameter $\gamma$ is nonnegative and the location parameter 
$\delta$ can be any real number with $-\infty < \delta < \infty$. The 
book by P. J. Nolan~\cite{Nolan:2020abc} details exhaustively the 
$\alpha=1$ special case and the ubiquitous nature of L\'evy-stable 
source distributions. It provides illustrations of the stable densities 
in the $S(\alpha, \beta, \gamma, \delta; 1)$ parameterization.

The asymmetric $\tau$-model~\cite{Csorgo:1990up} was utilized 
successfully by the L3 Collaboration~\cite{L3:2011kzb} to interpret 
two-jet data. Two of the experimentally testable predictions that are 
directly related to asymmetrical ($\beta\neq0$) source distributions 
were investigated. The first indication of such an asymmetry was related 
to a dip (an anticorrelated region) in the two-particle Bose-Einstein 
correlation function. Observance of such a dip in PHENIX data would be a 
strong experimental indication of the presence of an asymmetric source. 
However, such a dip is not observed in any of the investigated 
transverse-mass or centrality bins. Secondly, if the asymmetric 
$\tau$-model has relevance in the present analysis, the BEC function 
would depend on any relative-momentum component only through the 
invariant momentum $q_{\rm inv}$.  In this case, the BEC functions 
increase with decreasing values of $q_{\rm inv}$, even if $Q$ is kept 
constant. Such behavior is not observed here; on the contrary, the BEC 
function is seen to increase with decreasing $Q$, even if $q_{\rm inv}$ 
is kept at constant values. Therefore, a vanishing asymmetry parameter, 
$\beta = 0$, is assumed.

The correlation function is thus based on the assumption of the simplest 
case of univariate and symmetric ($\beta = 0$) stable distributions.  
The scale parameter $\gamma$ is replaced by the physical parameter 
$R$~\cite{Csorgo:2003uv}.  In the standardized notation of 
Nolan~\cite{Nolan:2020abc}, this corresponds to the $(\alpha, \beta{=}0, 
\gamma{=}R/2^{\frac{1}{\alpha}} , \delta; 1)$ convention. The 
Fourier-transformed source-density distribution has a simple form, 
$f(q)=\exp(iq \delta - \frac{1}{2}|q R|^{\alpha})$ and its modulus 
square leads to the simple form of the BEC function of 
Eq.~\eqref{eq:BEC-stable}. Note that the correlation function does not 
depend on $\delta$.

The relationship with the Gaussian source distribution is also apparent 
in Eq.~\eqref{eq:BEC-stable} as it corresponds to the $\alpha = 2$ 
special case. As detailed in Ref.~\cite{Csorgo:2003uv}, the physical 
scale parameter $R$ corresponds only in this $\alpha = 2$ special case 
to the root-mean-square of the source. For all $0<\alpha<2$, the 
root-mean-square is divergent, as is well known for the Cauchy or 
Lorentzian special case ($\alpha=1$). In the $\alpha = 1$ case the 
physical L\'evy scale parameter $R$ corresponds to the half width at 
half maximum (HWHM) of the source distribution. In fact for all values 
of $\alpha$, $R$ is proportional to the HWHM, the constant of 
proportionality depending on $\alpha$.  Another notable property of the 
L\'evy stable source distributions is that in the $\alpha < 1$ cases, 
even the first moment of the source distribution is divergent. In high 
energy particle and nuclear physics, the $R$ values correspond to a few 
femtometers~\cite{L3:2011kzb,PHENIX:2017ino}.

\subsubsection{The Coulomb interaction \label{sec:CCcoul} }

As in Ref.~\cite{PHENIX:2017ino}, the Coulomb final-state interaction is 
characterized using the Sinyukov-Bowler 
method~\cite{Akkelin:1995gh,1991PhLB27069B}. This method corresponds to 
the integration of the two-particle Coulomb wave 
function~\cite{Alt:1999cs} for a core-halo type of particle-emitting 
source~\cite{Csorgo:1994in,Bolz:1992hc}. However, such a Coulomb 
wave-function integration cannot be performed analytically in the case 
of a L\'evy source. Hence, numerical approaches are needed. The previous 
L\'evy BEC analysis used an iterative method based on a numerical table 
which contains the values of the integral for a range of values in the 
parameters and in the variable. The details can be found in 
Ref.~\cite{PHENIX:2017ino}. In this paper a 
parameterization~\cite{Csanad:2019cns,Csanad:2019lkp} is based on the 
aforementioned numerical table, which is considerably faster.

The momentum difference variable of the Coulomb correction is the 
invariant four-momentum difference $q_{\rm inv}$ rather than the 
variable $Q$, which is used in the present analysis. Neglecting the 
difference between the two variables could introduce a systematic 
uncertainty of $\approx$5\%~\cite{PHENIX:2017ino}. In the present 
analysis, the difference is determined by measuring the actual pair 
distribution in both $Q$ and $q_{\rm inv}$, $A(Q,q_{\rm inv})$. Using 
this two-dimensional distribution, the Coulomb correction is 
incorporated with a weighted average.

The final form of the fitting function is then
\begin{widetext}
\begin{align}
    C_2(Q;\lambda,R,\alpha,N,\varepsilon) &= 1 -\lambda + \lambda \: C_2^{(0)}(Q;R,\alpha,N,\varepsilon) \: w(Q;R,\alpha) \nonumber \\
    {\rm with} \hspace{1em} C_2^{(0)}(Q;R,\alpha,N,\varepsilon) &= (1+\exp(-R^\alpha Q^\alpha)) N (1+\varepsilon Q) \nonumber \\
    {\rm and} \hspace{1em} w(Q;R,\alpha) &= \frac{\sum_k A(Q,q_{{\rm inv,}k})K(q_{{\rm inv,}k};R,\alpha)}{\sum_k A(Q,q_{{\rm inv,}k})}
\label{eq:C2_full_definition}
\end{align}
\end{widetext}

where $K(q_{{\rm inv,}k};R,\alpha)$ is the Coulomb correction given by 
the parameterization~\cite{Csanad:2019cns,Csanad:2019lkp}. In the 
definition of the weight function $w(Q;R,\alpha)$, the index of 
summation $k$ runs over those bins in $q_{\rm inv}$ for a given value of 
$Q$, where the number of actual pairs after the two-track cuts is 
nonvanishing. This summation thus averages over the $q_{\rm 
inv}$-dependent Coulomb correction for a L\'evy-type source 
characterized by $R$ and $\alpha$, in $q_{\rm inv}$ bins only where 
$A(Q,q_{{\rm inv,}k})$ is nonzero. This method generalizes on Eqs. (2) 
and (3) of Ref.~\cite{PHENIX:2004yan} for a L\'evy-shaped Bose-Einstein 
correlation function and also accounts for the $q_\textmd{inv}$ 
dependence of the Coulomb correction. The large $Q$ behavior was found 
to be consistent with a linear function, which is characterized with the 
functional form of $N (1+\varepsilon Q)$.

\section{Systematic uncertainties}
\label{sec:systematics}

Nine sources of systematic uncertainties are investigated: the single 
track and the pair cuts, the choice of the arm of the PHENIX detector 
(because the arrangement is not symmetric), the fit range of the 
correlation functions, and the Coulomb-correction method. The systematic 
uncertainties of the results are estimated by varying one setting at a 
time, while keeping the others at their default values. For the cuts 
this means applying stricter or looser criteria. This is the same 
approach as used in Ref.~\cite{PHENIX:2017ino}, which see for details.  
For the pair cuts the values listed in Table~\ref{tab:paircuts-alt} are 
used rather than those of Ref.~\cite{PHENIX:2017ino}. For the choice of 
the arm of PHENIX the variation uses only one arm rather than both. The 
sensitivity of the results to the fit range was investigated by adding 
or leaving out one bin from the beginning ($Q_{\rm min}$) or the end of 
the fit range ($Q_{\rm max}$). The results do not vary significantly 
with the variation of $Q_{\rm max}$, therefore this source does not 
contribute to the systematic uncertainty.

As a default Coulomb-correction method, the parameterization detailed in 
Refs.~\cite{Csanad:2019cns,Csanad:2019lkp} is used.  A recent 
theoretical investigation~\cite{Kurgyis:2020vbz} suggested several 
alternative methods.  The systematic uncertainty of the Coulomb 
correction is taken as the difference between our approach and the most 
realistic variant of the Coulomb corrections mentioned in Section 3.2 of 
Ref.~\cite{Kurgyis:2020vbz}, namely the 6$^{\rm th}$ one in the 
enumeration.  The total systematic uncertainty is estimated with a 
standard statistical approach.  The individual contributions are summed 
quadratically, while both the statistical uncertainties $\sigma_{\rm 
stat}$ and the correlations between the uncertainties (denoted by 
$\rho$) are considered.  Thus, the final systematic uncertainty 
($\sigma_{\rm syst}$) is expressed for a parameter, $p$ (default cut 
denoted by $p_{\rm def}$ and the $i$th alternative cut by $p_{{\rm 
cut},i}$), with the following form: \begin{widetext}
\begin{align}
    \sigma_{\rm syst,tot}^2 = \sum_i \left[ (p_{\rm def}-p_{{\rm cut},i})^2-\sigma^2_{\rm stat}(p_{\rm def})-\sigma^2_{\rm stat}(p_{{\rm cut},i})+2\rho_i \sigma_{\rm stat}(p_{\rm def})\sigma_{\rm stat}(p_{{\rm cut},i})\right].
\end{align}
\end{widetext}
The $\rho_i$ correlation between the uncertainties can be estimated with 
a data-driven method by measuring the numbers of pairs yielded with the 
different settings
\begin{align}
    \rho_i =
    \begin{cases}
      \sqrt{\frac{N_{\rm def}}{N_{{\rm cut},i}}} & {\rm if} \hspace{0.5cm} N_{\rm def} < N_{{\rm cut},i} \\
      \sqrt{\frac{N_{{\rm cut},i}}{N_{\rm def}}} & {\rm if} \hspace{0.5cm} N_{{\rm cut},i} < N_{\rm def} .\\
    \end{cases}
\end{align}
The $\rho_i$ correlation coefficients typically are near unity, 
except for the third-pad-chamber matching cut and the arm settings, 
for which $\rho_i {\approx}0.6$.


\section{Centrality and transverse-mass-dependent results}
\label{sec:results}

The transverse-mass ($\mT$) and centrality dependence (expressed in 
terms of the average number of participants, $\Npart$) of the L\'evy 
parameters analyzed and are investigated along with their theoretically 
motivated combinations. The transverse mass is defined as 
$\mT~{=}~\sqrt{m_\pi^2 + K_T^2}$, where $K_T=\sqrt{K_x^2 + K_y^2}$ is 
the transverse component of the average momentum of the pair 
$K=0.5(p_1+p_2)$ and $m_\pi$ is the pion mass. The number of 
participants $\Npart$ was determined via Glauber-model calculations 
based on Ref.~\cite{Miller:2007ri}. The centrality dependencies of the 
L\'evy parameters are characterized by theoretically or empirically 
motivated functions. The fits are found to represent the data in each of 
the investigated centrality-class and transverse-mass bins, the 
confidence levels are in the statistically acceptable 
0.1\%$\leq$CL$\leq$99.9\% region. The fitted correlation functions 
are very similar to our results obtained in the 0\%--30\% centrality 
class.  For example, a L\'evy fit to our Bose-Einstein correlation data 
was published in Ref.~\cite{PHENIX:2017ino}.


\subsection{The centrality and transverse-mass dependence of the fit 
parameters}
\label{sec:mt_dep}

The dependence of the physical parameters $\lambda$, $R$, and $\alpha$ 
on centrality and transverse mass, $\mT$ is determined. The parameters 
of the linear background in Eq. \eqref{eq:C2_full_definition} are found 
to be $N \approx 1$ and $\varepsilon \approx 0$. In particular, the 
maximum of the modulus of the coefficient of linearity is found to be 
$\max(|\epsilon|)=0.085$ GeV$^{-1}$ and the average value is 
$\langle|\epsilon|\rangle=0.021\pm0.001$ GeV$^{-1}$. The overall 
normalization coefficient $N$ has a maximal deviation from unity of 
$\max(|N-1|)=0.015$, while its average deviation from unity is 
$\langle|N-1|\rangle=0.0050\pm0.0001$.

\begin{figure}
    \includegraphics[width=1.0\linewidth]{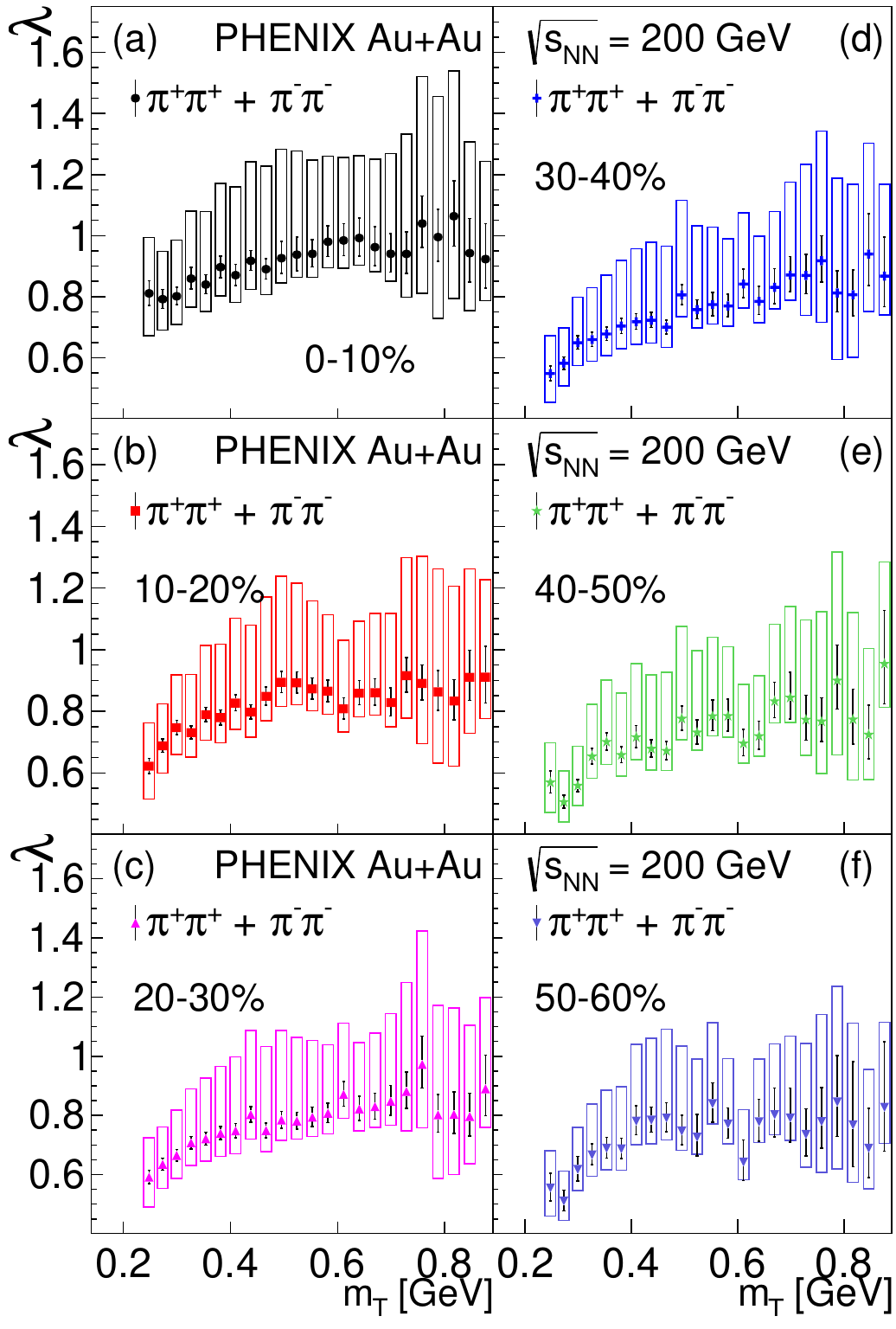}
    \caption{The transverse-mass dependence of the correlation-strength 
parameter $\lambda$ in six centrality bins obtained from L\'evy fits with 
Eq.~\eqref{eq:C2_full_definition}. The central values are shown with 
dots, statistical uncertainties are indicated by vertical lines, while 
boxes are used to illustrate the systematic uncertainties.}
    \label{fig:lambda_mt}
\end{figure}

\begin{figure}
    \includegraphics[width=1.0\linewidth]{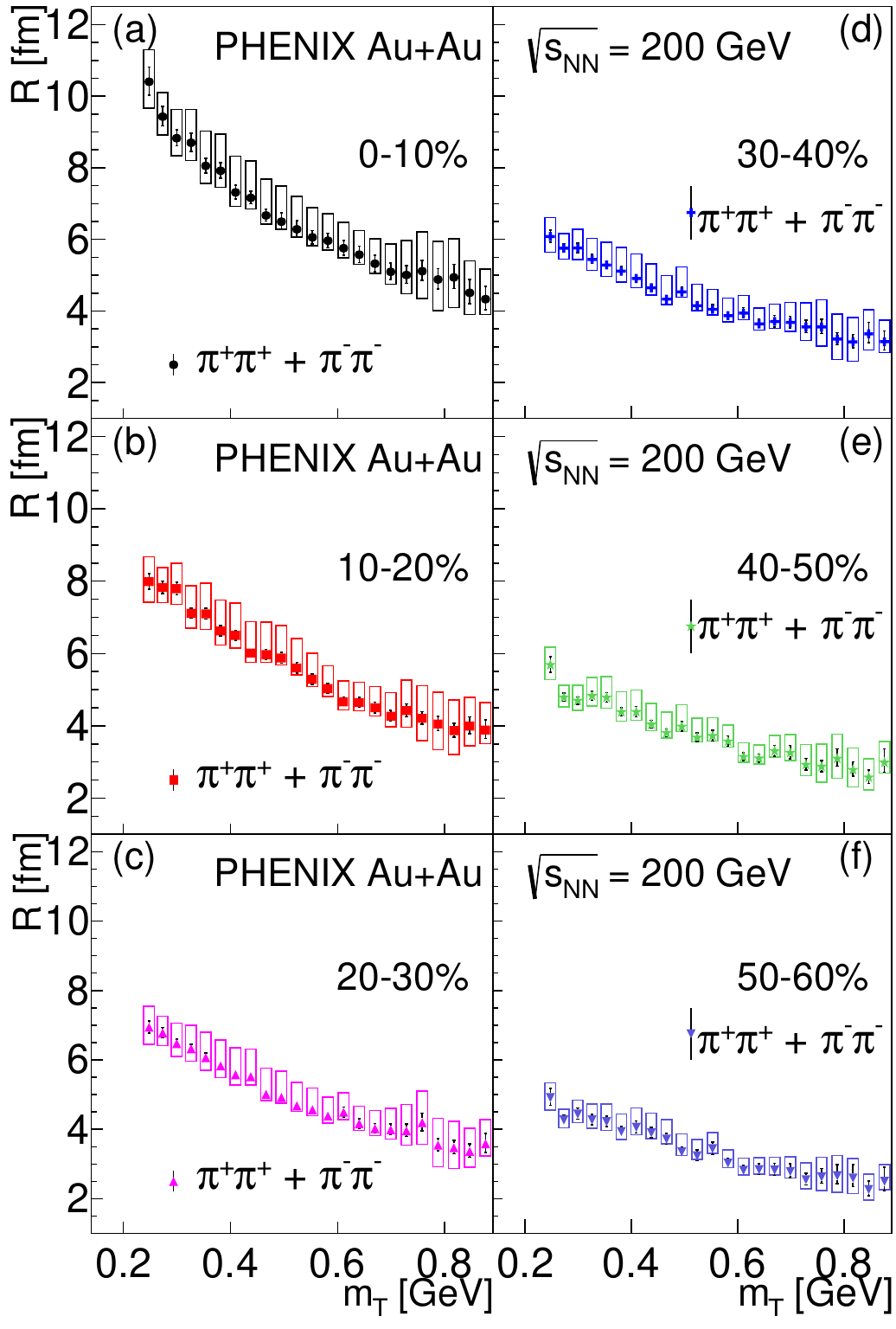}
    \caption{The transverse-mass dependence of the L\'evy-scale parameter 
$R$ in six centrality bins obtained from L\'evy fits with 
Eq.~\eqref{eq:C2_full_definition}. The central values are shown with 
dots, statistical uncertainties are indicated by vertical black lines, 
while boxes are used to illustrate the systematic uncertainties.}
    \label{fig:R_mt}
\end{figure}

\begin{figure}
    \includegraphics[width=1.0\linewidth]{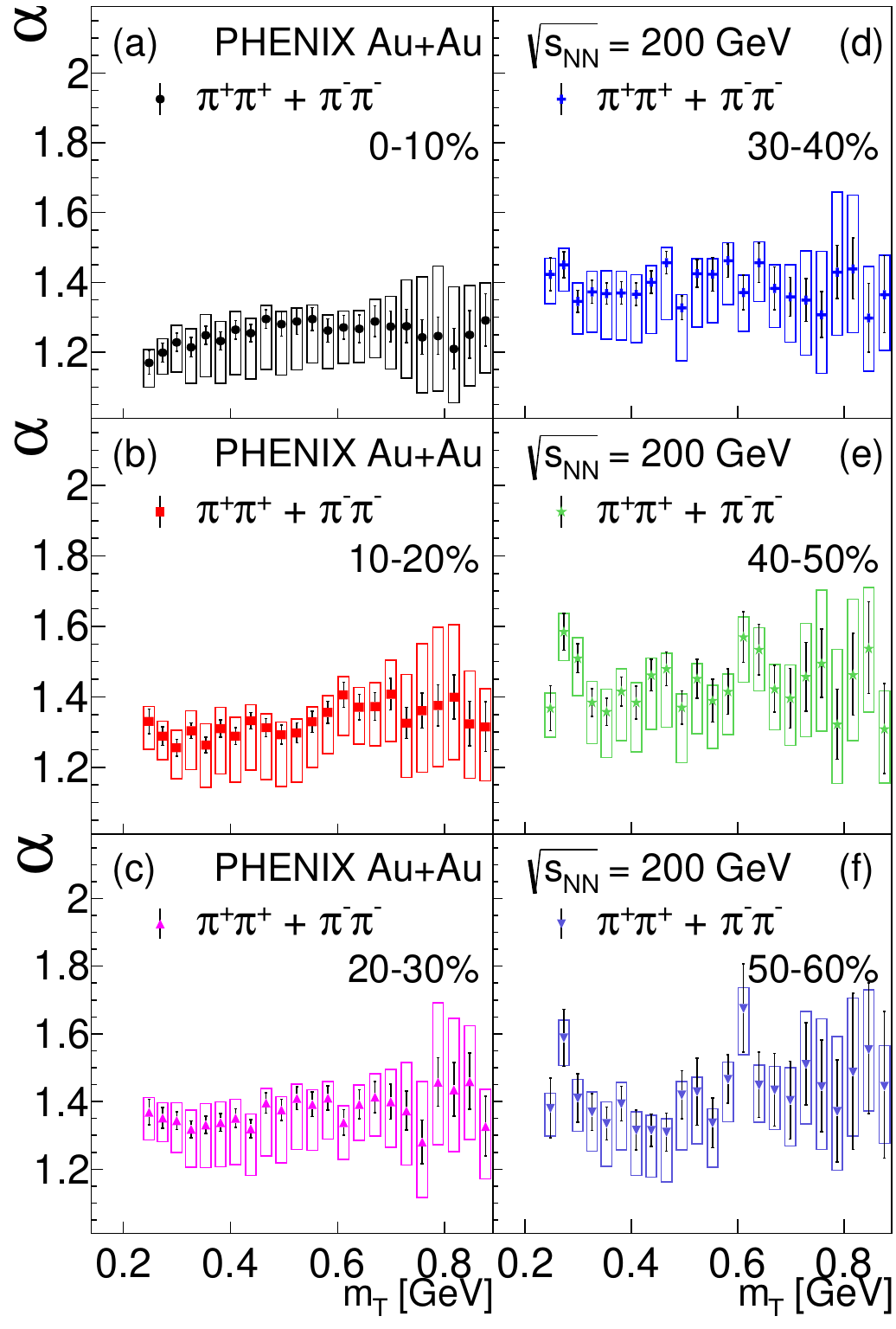}
    \caption{The transverse-mass dependence of the L\'evy-index of 
stability parameter $\alpha$, shown in six centrality bins obtained from 
L\'evy fits with Eq.~\eqref{eq:C2_full_definition}. The central values 
are shown with dots, statistical uncertainties are indicated by vertical 
black lines, while boxes are used to illustrate the systematic 
uncertainties.}
    \label{fig:alpha_mt}
\end{figure}

\begin{figure}
    \includegraphics[width=1.0\linewidth]{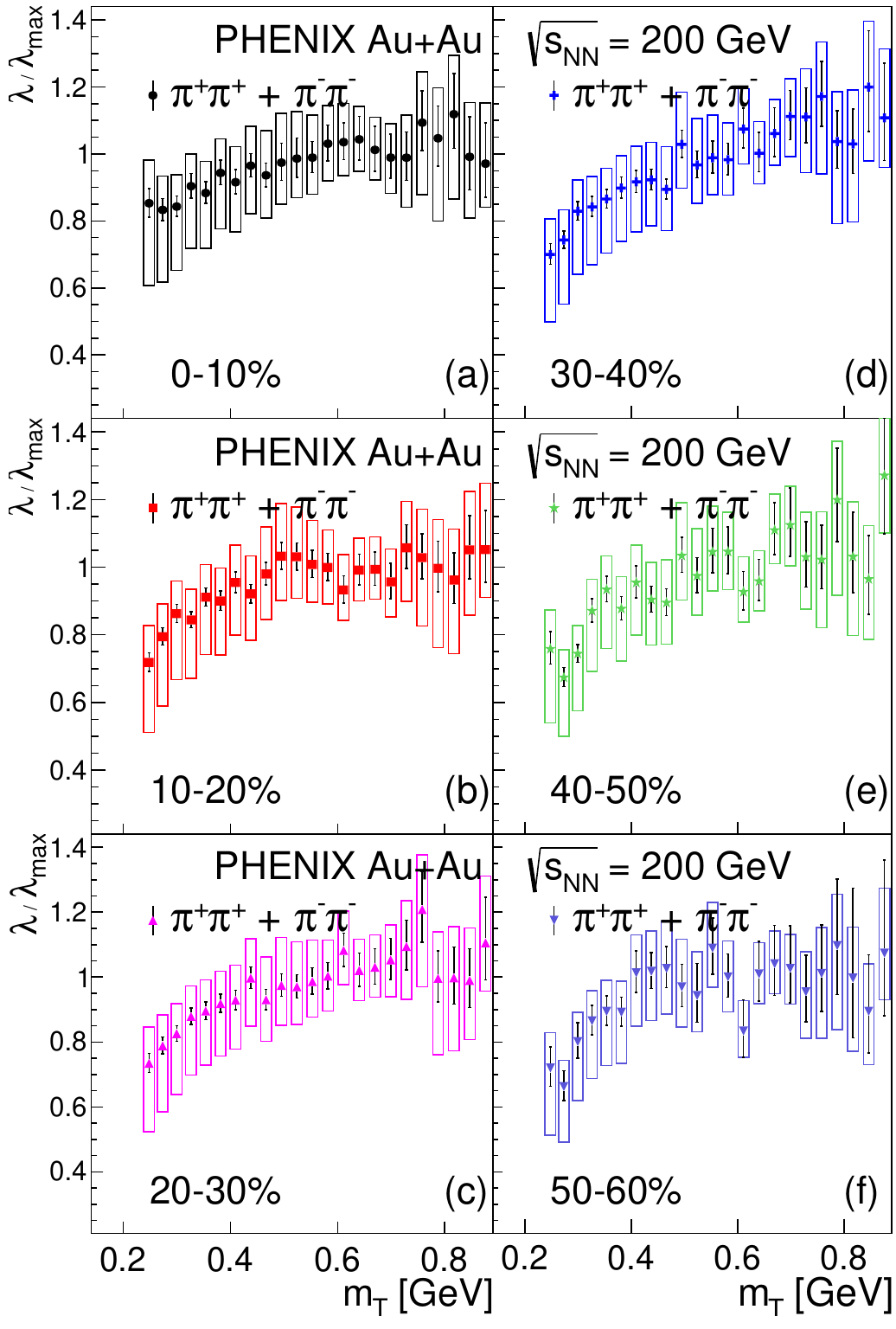}
    \caption{The transverse-mass dependence of the normalized 
correlation-strength parameter $\lambda/\lambda_{\rm max}$ for six 
centrality intervals obtained by rescaling Fig.~\ref{fig:lambda_mt} 
with a centrality-dependent $\lambda_\textmd{max}$ defined as the 
average value of $\lambda(\mT)$ in the $0.45 \leq \mT \leq 0.9$ GeV 
interval. The central values are shown with dots, statistical 
uncertainties are indicated by vertical black lines, while boxes are 
used to illustrate the systematic uncertainties. For each centrality bin 
the data are fitted with the Gaussian function of 
Eq.~\eqref{eq:1minusG}. This parameterization can describe the data and 
is discussed in detail in Section~\ref{sec:npart_dep}. The fit 
parameters with the statistic and systematic uncertainties are shown in 
Fig.~\ref{fig:Hs_npart}.}
    \label{fig:lambdamax_mt}
\end{figure}

\begin{figure}
    \includegraphics[width=1.0\linewidth]{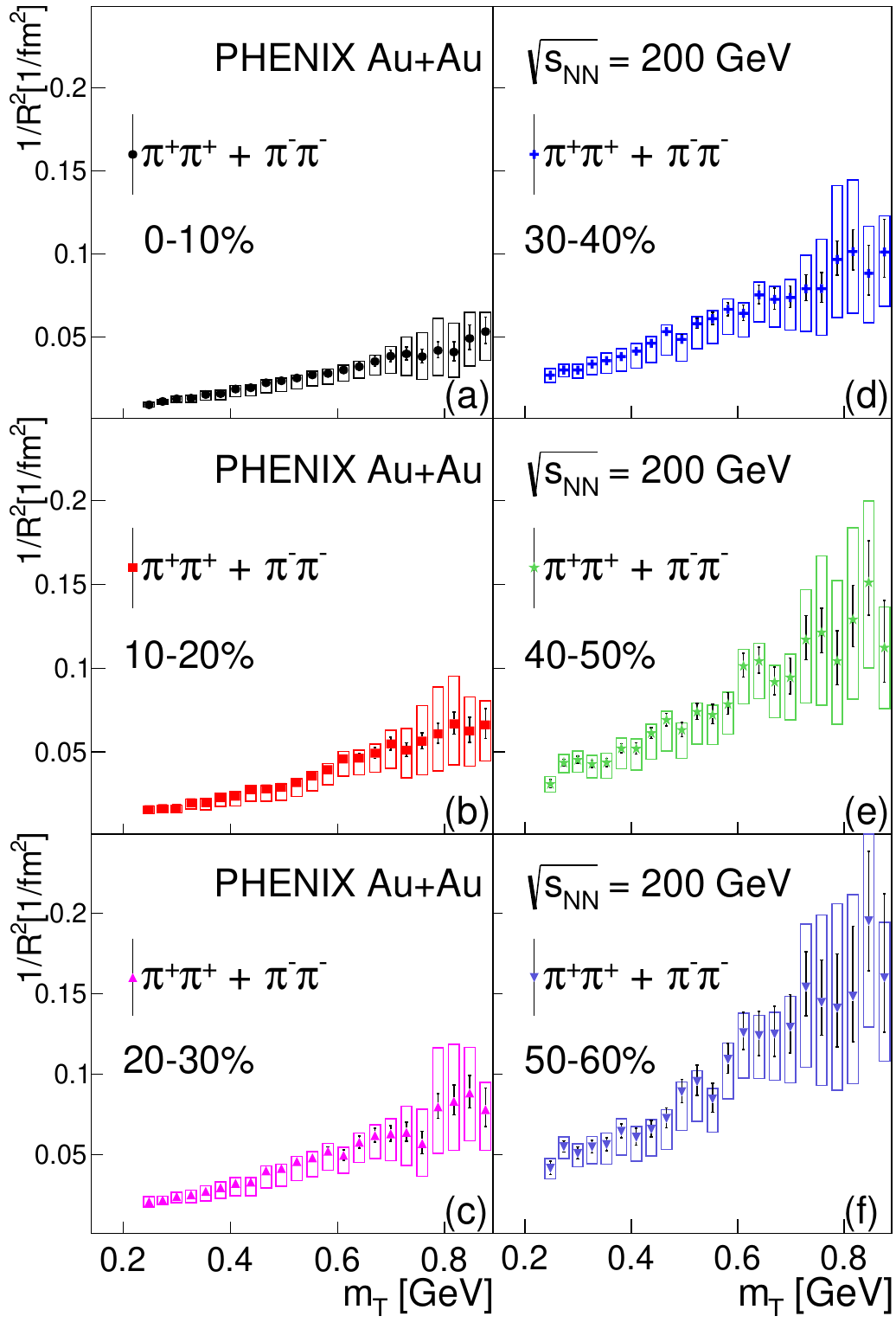}
    \caption{The transverse-mass dependence of the inverse square of the 
L\'evy-scale parameter $R$, shown in six centrality bins. The central 
values are shown with dots, statistical uncertainties are indicated by 
vertical black lines, while boxes are used to illustrate the systematic 
uncertainties. For each centrality bin the data are fitted with the 
linear function of Eq.~\eqref{eq:1overR2}. This parameterization can 
describe the data and is discussed in detail in 
Section~\ref{sec:npart_dep}. The fit parameters with the statistic and 
systematic uncertainties are shown in Fig.~\ref{fig:AB_npart}.}
    \label{fig:1overR2_mt}
\end{figure}

The results of the fits for the 6 centrality and 23 $\mT$ bins are shown 
in Figs.~\ref{fig:lambda_mt}--\ref{fig:alpha_mt}. The intercept or the 
correlation-strength parameter $\lambda$ has a clearly observable 
suppression below $\mT \leq 0.5$ GeV average pair transverse mass. Above 
that it saturates at a centrality-dependent value.  This low-$\mT$ 
suppression, which is characterized in more detail in 
Section~\ref{sec:npart_dep} is observed in all centrality bins; however, 
in the most-peripheral (50\%--60\%) centrality class, this observation 
starts to be limited by statistics. This is one of the reasons that no 
data are shown for the most peripheral, 60\%--95\% centrality class.

The saturation value of the intercept parameter could depend on 
background processes which are not in the scope of the present analysis. 
As was shown in Ref.~\cite{Csanad:2005nr}, the value of the intercept 
parameter is lower for Gaussian, intermediate for an Edgeworth 
expansion~\cite{Csorgo:2000pf}, and higher for an exponential shape, 
which seems to be a systematic effect that depends on how the functional 
form of the Bose-Einstein correlation function extrapolates the data to 
the $Q = 0$ limit. This systematic effect, an overall vertical 
uncertainty, can however be removed by normalizing to the saturation 
value of $\lambda(\mT)$ at large $\mT$, following 
Refs.~\cite{Csanad:2005nr,PHENIX:2017ino}.  The saturation value is 
taken as the average value of $\lambda(\mT)$ in the interval 
0.45~GeV~$\leq{\mT}\leq$~0.9~GeV and is denoted by $\lambda_{\rm max}$. 
This centrality-independent range is the same for each of the six centrality 
classes considered in this manuscript. 
However, from the similar range 
considered in Ref.~\cite{PHENIX:2017ino} for the 0\%--30\% centrality 
selection, this range is slightly modified due to the different 
centrality classes considered. This range modification shifts the 
central values of $\lambda_\textmd{max}$ slightly, but the modification 
is within one standard deviation, which is within the uncertainties 
given in Ref.~\cite{PHENIX:2017ino}. However, the value of 
$\lambda_{\rm max}$ does depend on centrality.  The resulting 
$\lambda(\mT)/\lambda_{\rm max}$ ratio is shown in 
Fig.~\ref{fig:lambdamax_mt}.

Radial-flow effects are known to be strongly centrality dependent, and 
hence are expected to significantly influence both $\lambda(\mT)$ and 
$\lambda_{\rm max}$. Such an expectation is shown for example in 
Ref.~\cite{Vance:1998wd}. If there is large radial flow, the decay 
products of $\etaprime$ are concentrated at high $\mT$ (compensating 
$\pT$ sharing between the several daughter pions, resulting in flat 
$\lambda(\mT)/\lambda_{\rm max}(\mT)$), while if the radial flow is low 
(such as in the peripheral event classes), then these decay products 
accumulate at low $\mT$ (resulting in a dip in 
$\lambda(\mT)/\lambda_{\rm max}$ at low $m_T$). The two effects, the 
$\pT$ increasing radial flow and the $\pT$ sharing between the several 
pions from the $\etaprime \rightarrow \eta + \pi^+ +\pi^- \rightarrow 
(\pi^0 + \pi^+ +\pi^-) + \pi^+ +\pi^- $ decay chain compensate one 
another at transverse velocity $\langle u_T\rangle {\approx}0.5$, as 
shown in Ref.~\cite{Vance:1998wd}. This expectation is cross-checked 
with the Monte-Carlo simulations detailed in the Appendix, where the 
simulation-based centrality dependence of $\lambda(\mT)/\lambda_{\rm 
max}$ (without any in-medium $\etaprime$ mass modification) is also 
shown.

In Fig.~\ref{fig:lambdamax_mt}, an approximate centrality independence 
of the characteristics of the suppression can be qualitatively observed. 
To test this observation quantitatively, a $\chi^2$ test was employed to 
obtain conservative values of the CL using only statistical 
uncertainties. For each of the 15 possible pairs of centrality classes, 
the $\chi^2$ is calculated under the hypothesis that their 
$\lambda/\lambda_{\rm max}$ distributions are identical. The resulting 
CL values are all in the range 0.15\%--89.5\%, which does not reject the 
hypothesis.

The experimental result on the centrality independence of 
$\lambda(\mT)/\lambda_{\rm max}$ scaling is an unexpected, rather 
surprising observation. One possible explanation for such a scaling 
behavior is given in Section~\ref{sec:simulations}, based on the 
Monte-Carlo simulations detailed in the Appendix.  A quantitative 
analysis of these $\lambda(\mT)/\lambda_{\rm max}$ measurements is 
presented in Section~\ref{sec:npart_dep}.

The L\'evy-scale parameter, introduced in high-energy particle and 
nuclear physics as the physical-scale parameter $R$ in 
Ref.~\cite{Csorgo:2003uv}, decreases with $\mT$ for all centrality bins, 
as can be seen in Fig.~\ref{fig:R_mt}. Analytic hydrodynamic 
calculations~\cite{Akkelin:1995gh,Csorgo:1997us,Chapman:1994ax} predict 
that in the $\alpha=2$ special case, $R^{-2}(\mT)$ depends linearly on 
$\mT$, which can be parameterized as
\begin{align}
    \frac{1}{R^2} = A m_T + B
    \label{eq:1overR2}
\end{align}
as shown in Fig.~\ref{fig:1overR2_mt}. The centrality dependence of the 
slope parameter $A$ and the intercept parameter $B$ is presented in 
Section~\ref{sec:npart_dep}.

Figure~\ref{fig:alpha_mt} shows the measured values of the L\'evy index 
$\alpha$. In each centrality bin $\alpha$ appears to be independent of 
$\mT$.  In each centrality class, the data of Fig.~\ref{fig:alpha_mt} 
can be well represented by the $\mT$-averaged values, $\alpha_0$. As 
indicated in Fig.~\ref{fig:a0_npart}, these values are found to depend 
on centrality. The measured values of $\alpha$ are significantly lower 
than the Gaussian case of $\alpha=2$, and are significantly higher than 
the exponential/Cauchy case of $\alpha=1$. The centrality dependence of 
$\avgalpha$ is discussed in Section~\ref{sec:npart_dep}.

\begin{figure}
    \includegraphics[width=1.0\linewidth]{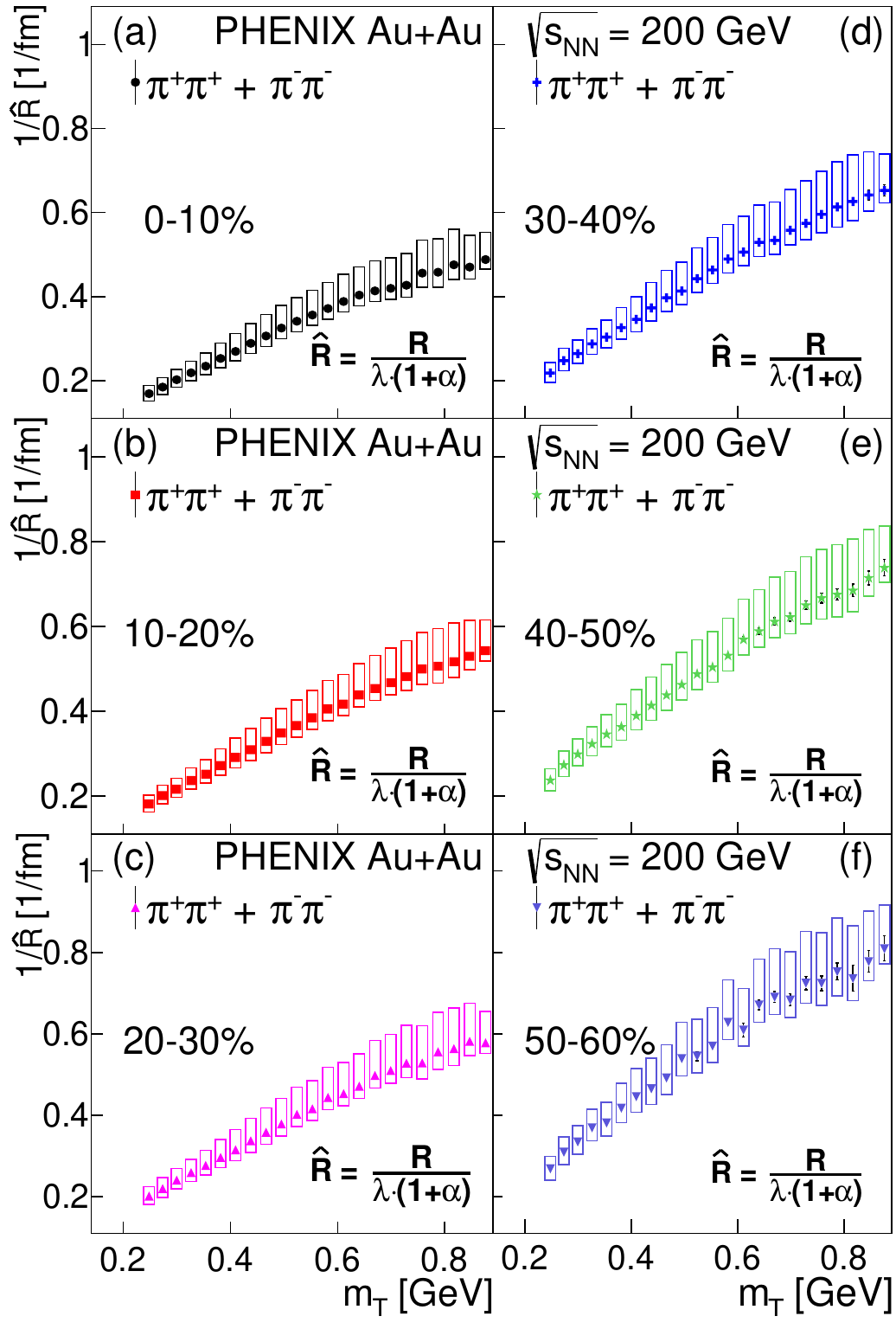}
    \caption{ The transverse-mass dependence of the $\widehat{R}$ 
parameter, shown in six centrality bins. The central values are shown 
with dots, statistical uncertainties are indicated by vertical black 
lines, while boxes are used to illustrate the systematic uncertainties. 
For each centrality bin the data are fitted with the linear function of 
Eq.~\eqref{eq:1overRhat}. This parameterization can describe the data 
and is discussed in detail in Section~\ref{sec:npart_dep}. The fit 
parameters with the statistic and systematic uncertainties are shown in 
Fig.~\ref{fig:ABhat_npart}.}
    \label{fig:Rhat_mt}
\end{figure}

Similar to earlier PHENIX studies in the {0\%--30}\% centrality-class 
range ~\cite{PHENIX:2017ino}, strong correlations are observed between 
the parameters $\lambda$, $R$, and $\alpha$. However, a specific 
combination of these three L\'evy parameters, namely
\begin{align}
    \frac{1}{\widehat{R}} = \frac{\lambda(1+\alpha)}{R},
\end{align}
shown in Fig.~\ref{fig:Rhat_mt}, seems to be less correlated with the 
direct fit of L\'evy parameters $\lambda$, $R$, $\alpha$ as compared to 
the correlations among them, which is very similar to the correlation 
plots published in Ref.~\cite{PHENIX:2017ino} for the 0\%--30\% 
centrality class.  The correlation coefficients between the parameters 
$\lambda$, $R$ and $\alpha$ were determined in the L\'evy fits and were 
considered when evaluating $\widehat{R}^{-1}$.

In Ref.~\cite{PHENIX:2017ino} $\widehat{R}^{-1}$ was found to depend 
approximately linearly on $m_T$ and was parameterized in the 
0\%--30\% centrality class as
\begin{align}
    \frac{1}{\widehat{R}} = \widehat{A} m_T + \widehat{B}.
    \label{eq:1overRhat}
\end{align}
Figure~\ref{fig:Rhat_mt} shows a more detailed centrality-dependent 
analysis. The centrality dependence of the $\widehat{A}$ slope parameter 
and $\widehat{B}$ intercept parameter is presented in 
Section~\ref{sec:npart_dep}. The linearity of $\widehat{R}^{-1}$ is 
broken at large $\mT$, which is likely due to the saturation of the 
$\lambda$ parameter in that range. As noted in 
Ref.~\cite{PHENIX:2017ino}, the understanding of the approximate scaling 
properties of $\widehat{R}$ is another currently unsolved theoretical 
challenge.

\subsection{The centrality dependence of the physical parameters}
\label{sec:npart_dep}

In this subsection, the centrality dependence of the physical parameters 
via theoretically or empirically motivated parameterizations are 
investigated.  The centrality dependence of the parameters of these 
parameterizations is discussed here as a function of $N_{\rm part}$. 
Quantitatively studying the suppression of $\lambda(\mT)/\lambda_{\rm 
max}$ at low $\mT$, uses fits of the phenomenological Gaussian 
parameterization, which was also used in Ref.~\cite{PHENIX:2017ino},
\begin{align}
\frac{\lambda}{\lambda_{\rm max}} = 1 - H \exp\left({-\frac{\mT^2-m_\pi^2}{2\sigma^2}}\right),
\label{eq:1minusG}
\end{align}
where the parameter $H=H(\Npart)$ measures the depth of the suppression 
and parameter $\sigma=\sigma(\Npart)$ measures the width of this 
suppression. In principle, both $H$ and $\sigma$ could have a centrality 
or $\Npart$ dependence.

\begin{figure}
    \includegraphics[width=1.0\linewidth]{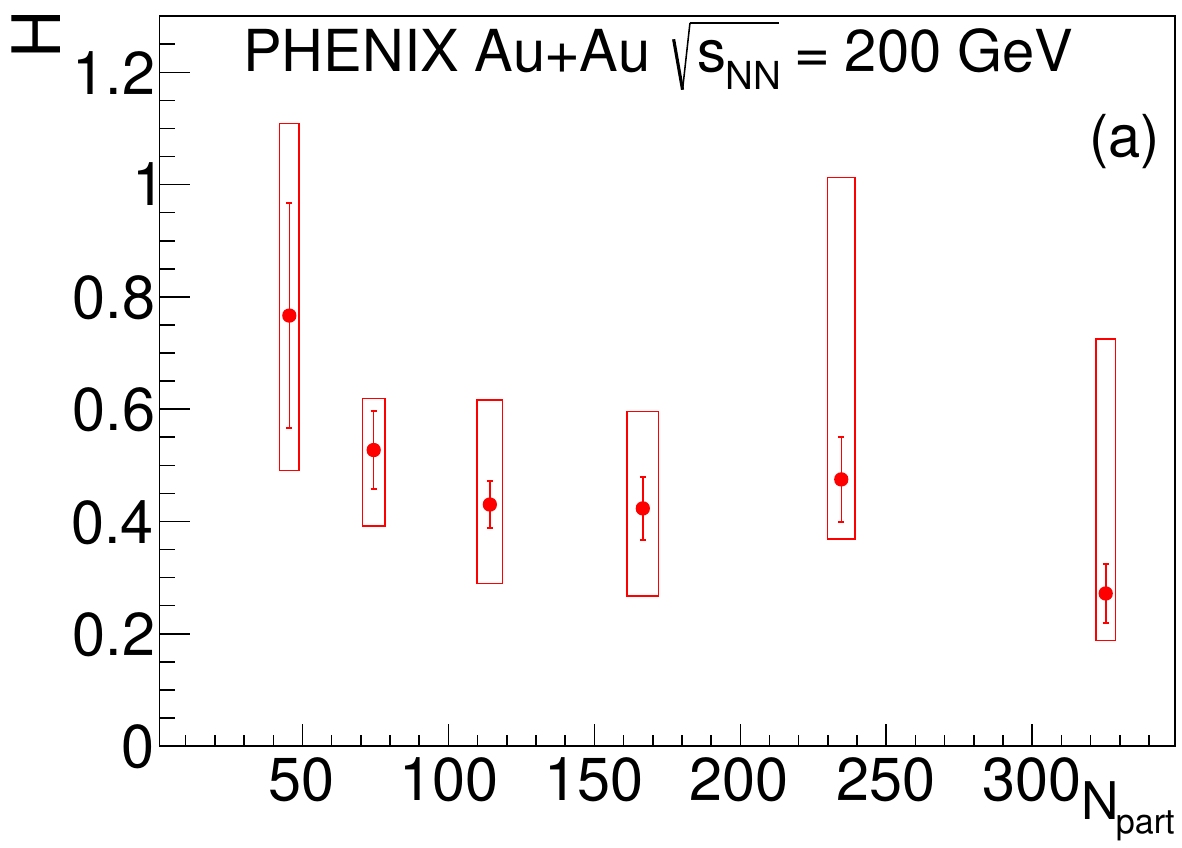}
    \includegraphics[width=1.0\linewidth]{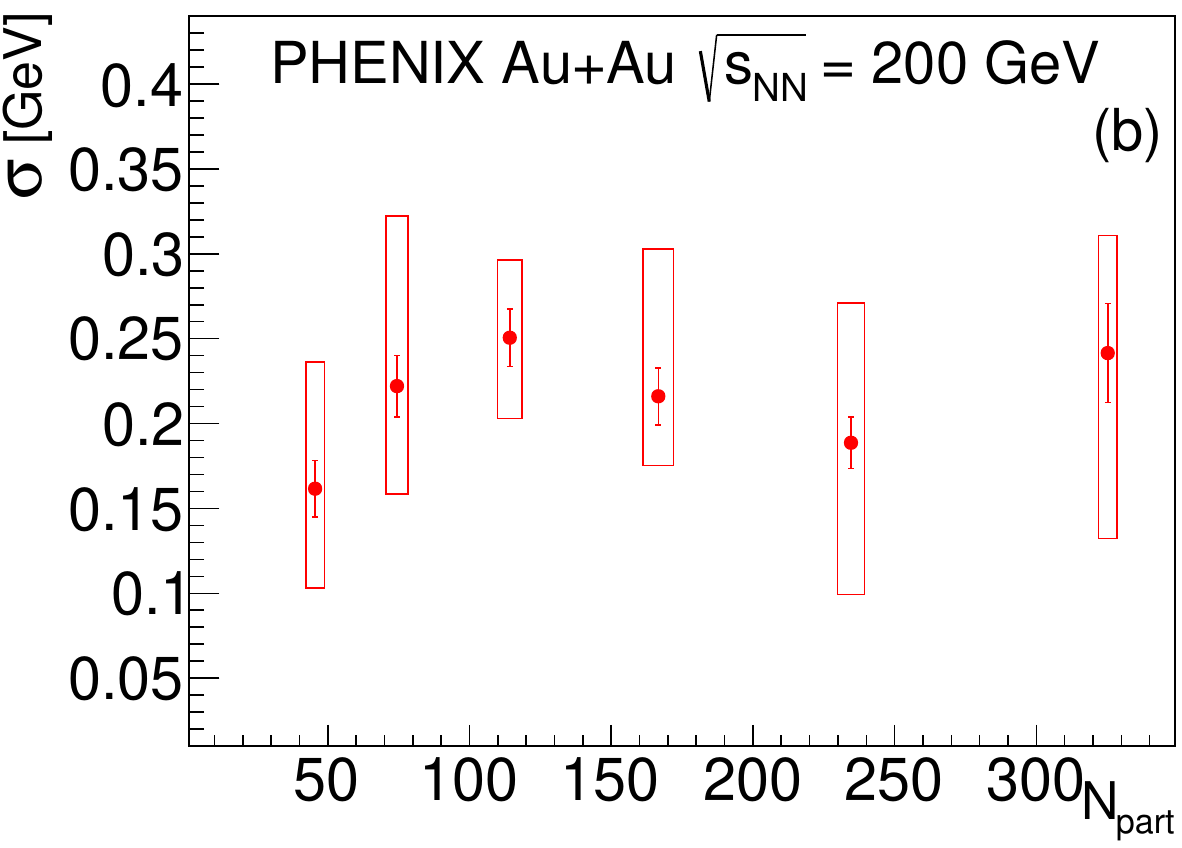}
    \caption{The two parameters that characterize the suppression of the 
$\lambda/\lambda_{\rm max}(\mT)$ and are defined in 
Eq.~\eqref{eq:1minusG}, (a) the magnitude ($H$), and (b) the width 
($\sigma$) are shown as functions of $\Npart$. The central values are 
shown with dots, statistical uncertainties are indicated by vertical [red] 
lines, while boxes are used to illustrate the systematic uncertainties. 
Both panels (a) and (b) are consistent with a centrality-independent 
constant values $H_0$, $\sigma_0$.}
    \label{fig:Hs_npart}
\end{figure}

From the experimentally observed, and rather surprising, 
centrality-independent scaling of $\lambda(\mT)/\lambda_{\rm max}$, the 
centrality independence of the $H$ and the $\sigma$ fit parameters is 
expected, as discussed in Section~\ref{sec:mt_dep}. Indeed, both 
obtained parameters are consistent with the hypothesis of centrality 
independent values $H_0$ and $\sigma_0$, as indicated in 
Fig.~\ref{fig:Hs_npart}.

\begin{figure}
    \includegraphics[width=1.0\linewidth]{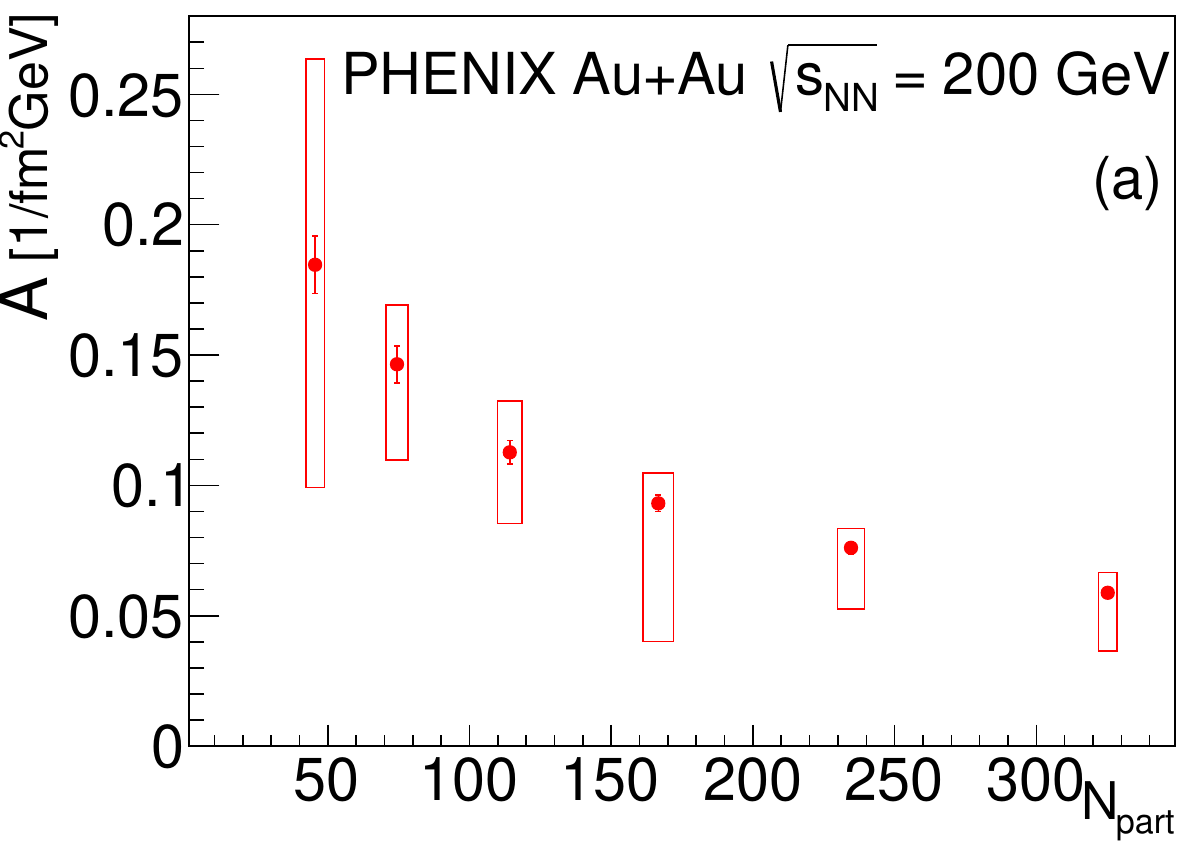}
    \includegraphics[width=1.0\linewidth]{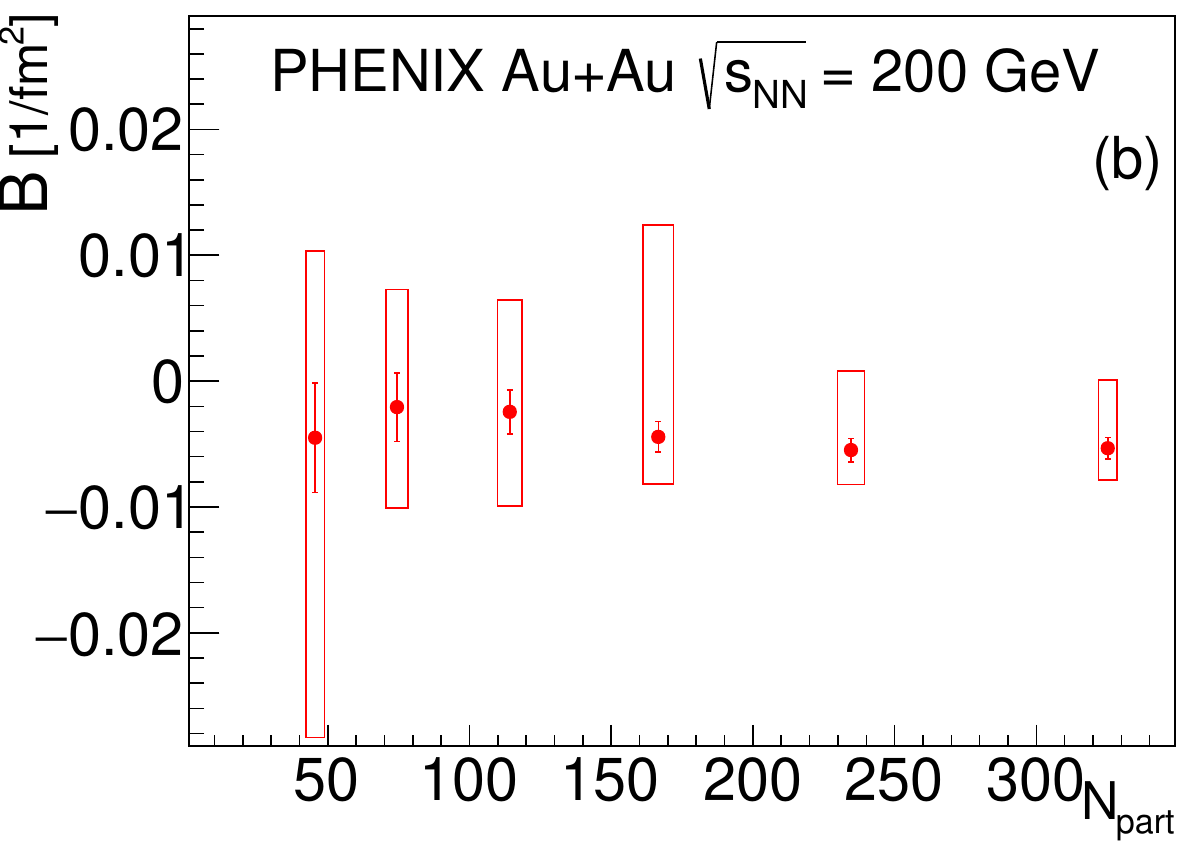}
    \caption{
    The two parameters of the affine linear fit to the inverse square of 
the L\'evy-scale parameter $1/R^2(\mT)$ are defined in 
Eq.\eqref{eq:1overR2}, (a) the slope parameter ($A$), and (b) the 
intercept parameter ($B$) are shown as functions of $\Npart$. The 
central values are shown with dots, statistical uncertainties are 
indicated by vertical [red] lines, while boxes are used to illustrate 
the systematic uncertainties.}
    \label{fig:AB_npart}
\end{figure}

The affine linearity of the inverse square of the L\'evy-scale parameter 
$R^{-2}(\mT)$, Eq.~\eqref{eq:1overR2}, is demonstrated in 
Fig.~\ref{fig:1overR2_mt}. The parameters of these fits as a function of 
$\Npart$ are shown in Fig.~\ref{fig:AB_npart}. The slope parameter $A$ 
decreases with $N_{\rm part}$. The intercept parameter $B$, which could 
be connected to the size of the source, is slightly negative within 
statistical uncertainties but is compatible with zero if systematic 
uncertainties are taken into account. Note that in hydrodynamical 
calculations with $\alpha = 2$, the parameter $B$ is dominated by the 
inverse of the squared geometrical size of the whole source: the smaller 
the value of B, the larger the geometrical size of the source, while 
parameter $A$ is dominated by radial flow 
effects~\cite{Csorgo:1995bi,Chapman:1994ax,Kincses:2022eqq}.

As mentioned in Section~{\ref{sec:mt_dep}, the affine linearity of 
$\widehat{R}^{-1}(\mT)$, Eq.~\eqref{eq:1overRhat} is investigated, as a 
function of $\Npart$.  The centrality dependence of the slope and 
intercept parameters of these fits are shown in 
Fig.~\ref{fig:ABhat_npart}. The parameter $\widehat{A}$ exhibits a 
decreasing trend with increasing centrality similar to that seen for $A$ 
in the parameterization of $R^{-2}$.  Within statistical uncertainty, 
$\widehat{B}$ is compatible with a small positive number but both 
statistical and systematic uncertainties are too large to draw a strong 
conclusion from the present data.

\begin{figure}
    \includegraphics[width=1.0\linewidth]{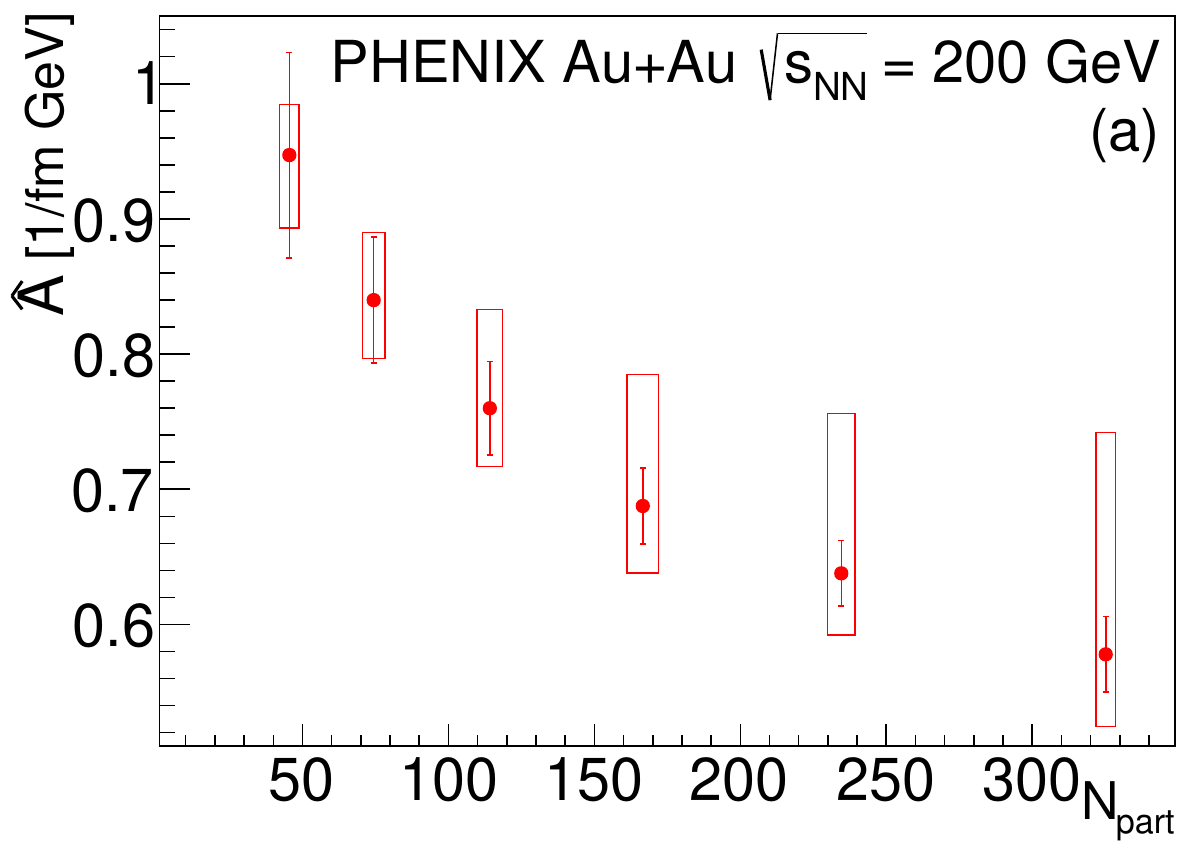}
    \includegraphics[width=1.0\linewidth]{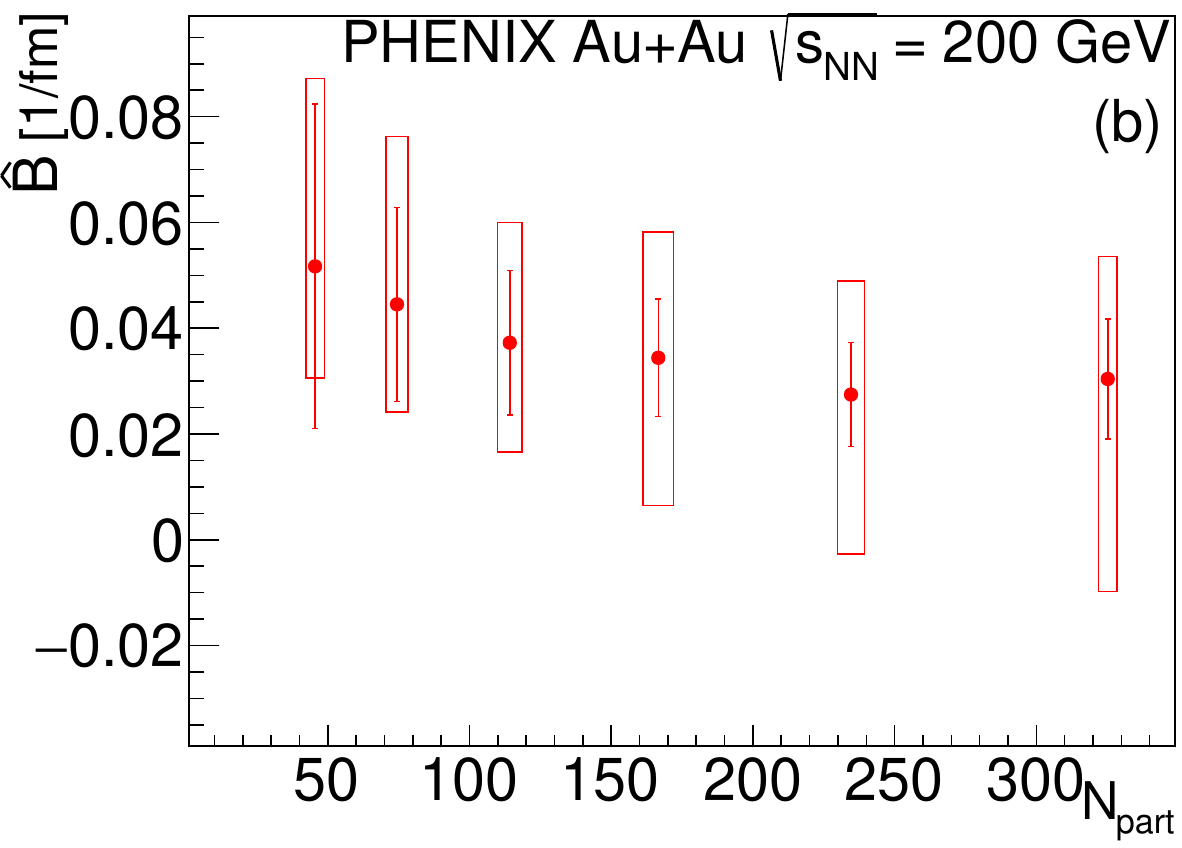}
    \caption{The two parameters of the affine linear fit to the inverse 
square of the L\'evy-scale parameter $1/\widehat{R}(\mT)$ are defined in 
Eq.\eqref{eq:1overRhat}, (a) the slope parameter ($\widehat{A}$), and 
(b) the intercept parameter ($\widehat{B}$) are shown as functions of 
$\Npart$. The central values are shown with dots, statistical 
uncertainties are indicated by vertical red lines, while boxes are used 
to illustrate the systematic uncertainties.}
    \label{fig:ABhat_npart}
\end{figure}

As shown in Fig.~\ref{fig:a0_npart}(a), because the L\'evy-exponent 
parameter $\alpha$ does not appear to depend on $\mT$ 
(see Fig.~\ref{fig:alpha_mt}), the $\mT$-averaged value, $\avgalpha$, can 
be used to determine the centrality dependence of $\alpha$.  A significant 
centrality dependence of $\avgalpha$ is observed. The lowest value of 
$\avgalpha$ occurs for the most central collisions. As the collisions 
become more peripheral $\avgalpha$ saturates around $\approx$1.4.

The parameter $R$ is investigated in several $\mT$ bins as a function of 
$\Npart^{1/3}$. The linearity of this dependence, shown in 
Fig.~\ref{fig:a0_npart}(b), suggests that the volume of the L\'evy 
source, which is proportional to $R^3$, is proportional to $\Npart$. 
More detailed studies, investigating the centrality dependence in terms 
of constituent quark participants along the lines of 
Ref.~\cite{PHENIX:2013ehw} would be desirable but go beyond the scope of 
the present analysis.

\begin{figure}
    \includegraphics[width=1.0\linewidth]{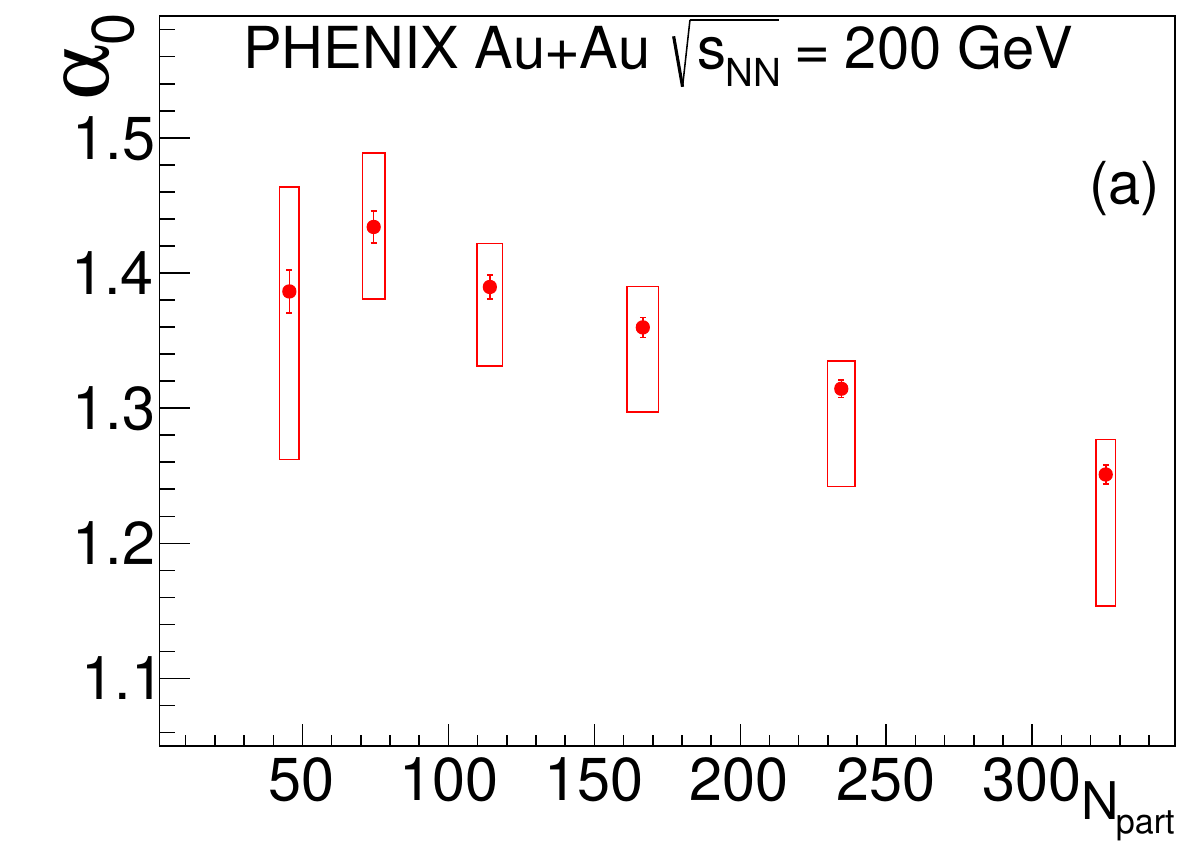}
    \includegraphics[width=1.0\linewidth]{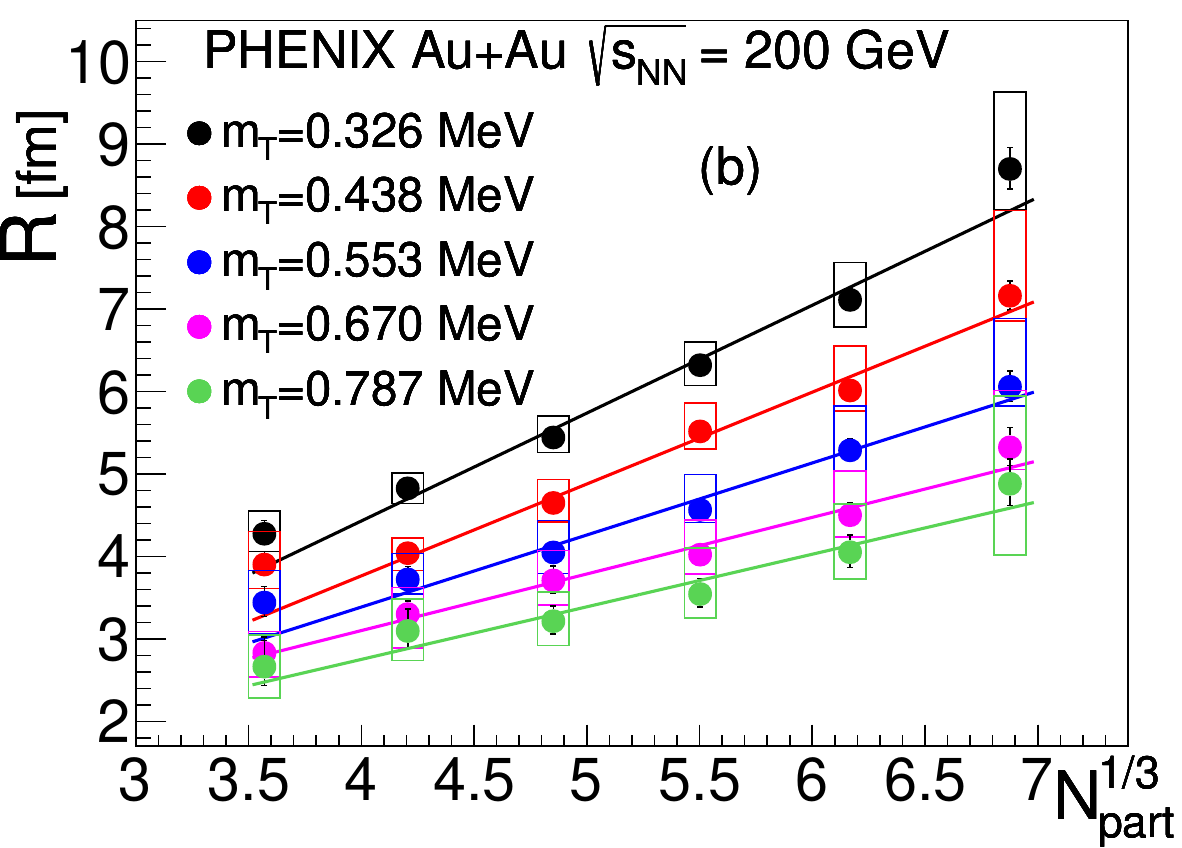}
    \caption{The centrality dependence of (a) $\avgalpha$, the 
$\mT$-averaged value of $\alpha$ as a function of $\Npart$ and (b) the 
L{\'e}vy scale parameter $R$ as a function of $\Npart^{1/3}$ in selected 
$\mT$ bins.}
    \label{fig:a0_npart}
\end{figure}


\section{Comparison with Monte-Carlo simulations}
\label{sec:simulations}

The low-$\mT$ decrease of the $\lambda(\mT)$ correlation-strength 
measurements~\cite{PHENIX:2004yan,STAR:2009fks,STAR:2004qya} of charged 
pions is found in Ref.~\cite{Vance:1998wd} to be an indirect signal of 
the in-medium mass reduction of the $\etaprime$ particles. The 
$\lambda(\mT)$ suppression at low transverse mass ($\mT$) has been 
investigated in Au$+$Au collisions at $\sqrt{s_{NN}}$= 200 GeV 
collisions by Monte-Carlo 
simulations~\cite{Csorgo:2009pa,Vertesi:2009wf}. These calculations were 
also compared to the results of the previous PHENIX analysis of 
0\%--30\% central Au$+$Au collisions at 
$\sqrt{s_{NN}}=200$~GeV~\cite{PHENIX:2017ino}.  Although the 
measurements are performed in pair-$\mT$ (cf.~Section~\ref{sec:results}) 
ranges while the simulations are in the single-particle transverse mass 
$\sqrt{m^2+p_T^2}$, this difference can be neglected as $\lambda$ is a 
characteristic parameter of the two-particle correlation function at 
vanishing values of relative momentum, where $K{=}p$ and hence the 
single-particle transverse mass is equal to the pair $\mT$. This 
property is also utilized in the core-halo picture, where the smoothness 
approximation is warranted and $\lambda$ is evaluated at 
$K{=}p$~\cite{Vance:1998wd}.

The comparison of Monte-Carlo resonance-model simulations of 
$\lambda(\mT)/\lambda_{\rm max}$ to the above presented data shows that, 
within systematic uncertainties, an in-medium mass drop of $\etaprime$ 
is not inconsistent with our measured data. Theoretically, the mass of 
the $\etaprime$ meson could be sensitive to the $U_A(1)$ symmetry 
restoration in hot and dense hadronic 
matter~\cite{Pisarski:1983ms,Kapusta:1995ww}. The key point being that 
compared to the other pseudoscalar mesons, the $\etaprime$ is 
anomalously heavy, $\approx$958~MeV, although the $\etaprime$ quark 
content is similar to that of the $\eta$ meson. The $\eta$ meson mass, 
$m_\eta\approx$548~MeV, is closer to the mass of the charged kaons, 
$\approx{494}$~MeV while it is 410~MeV lighter than the mass of 
the $\etaprime$ meson, $m_\etaprime\approx{958}$~MeV, although the 
$\eta$ and the $\etaprime$ mesons have the same quark content 
~\cite{Workman:2022ynf}. The extra mass is explained in the Standard 
Model in terms of the $U_A(1)$ anomaly that couples the mass of the 
$\eta'$ to the topological properties of the quantum-chromodynamics 
vacuum state. If at high temperatures the structure of this vacuum 
changes, the extra 410 MeV mass difference may vanish and the 
$\etaprime$ mesons may return to the mass scale of the other 
pseudoscalar mesons, with its mass becoming similar to that of the 
$\eta$~\cite{Kapusta:1995ww}.  As the $\etaprime$ meson leaves the hot 
and dense matter, it regains its anomalously large mass at the expense 
of its kinetic energy and consequently this effect modifies the 
spectrum. See below for a discussion of how Monte-Carlo calculations are 
used to investigate the effect on $\lambda(\mT)$ of this mass 
modification, along the lines of 
Ref.~\cite{PHENIX:2017ino,PHENIX:2017ino,Vance:1998wd,Csorgo:2009pa,Vertesi:2009wf}.

For each centrality class the SHAREv3 Monte-Carlo 
generator~\cite{Wheaton:2004qb,Petran:2013dva} is used to evaluate the 
fraction of those short- and long-lived resonances that are the most 
important sources of pions. These fractions are used in our simulation. 
The input parameters to SHARE correspond to the STAR chemical freeze-out 
fits to the available experimental data on particle 
yields~\cite{PHENIX:2003iij,STAR:2010avo,STAR:2008med,BRAHMS:2002efn,STAR:2004yym,Gaudichet:2003jr,Suire:2002pa} 
and are available in the columns labeled grand-canonical-ensemble yields 
(GREY) of Table VIII in Ref.~\cite{STAR:2017sal}. The results from SHARE 
serve as inputs to our simulations.

The two most important parameters in the simulations are the in-medium 
mass of $\etaprime$ (denoted by $m^*_{\etaprime}$) and the effective 
temperature of the $\etaprime$ condensate, the inverse-slope parameter 
(denoted by $B_{\etaprime}^{-1}$). The in-medium mass drives the depth, 
while the inverse slope controls the steepness of the ``dip'' of the 
$\lambda(\mT)$ function. These two parameters are considered as fit 
parameters for the Monte-Carlo simulations. Further details of the 
simulation and the estimation of its systematic uncertainties are given 
in the Appendix and in 
Refs.~\cite{PHENIX:2017ino,Vance:1998wd,Csorgo:2009pa,Vertesi:2009wf}.

The $\chi^2$ scans are implemented by taking small steps in the 
in-medium $\etaprime$ mass and the inverse-slope parameter 
$B_{\etaprime}^{-1}$, comparing the results to the data.  From these 
scans are determined the optimal (minimum $\chi^2$) values and their 
uncertainties. The Appendix gives details of the simulation and the 
estimation of uncertainties.

Figure~\ref{fig:best-fits} compares the data to both the optimal fits 
and the case that lacks the in-medium mass modification of the 
$\etaprime$ meson. For $\mT\ageq 500$ MeV there is no significant 
difference between the optimal fit and the assumption of no modification 
case. At low $\mT$ no modification is strongly disfavored.

\begin{figure}
    \includegraphics[width=1.0\linewidth]{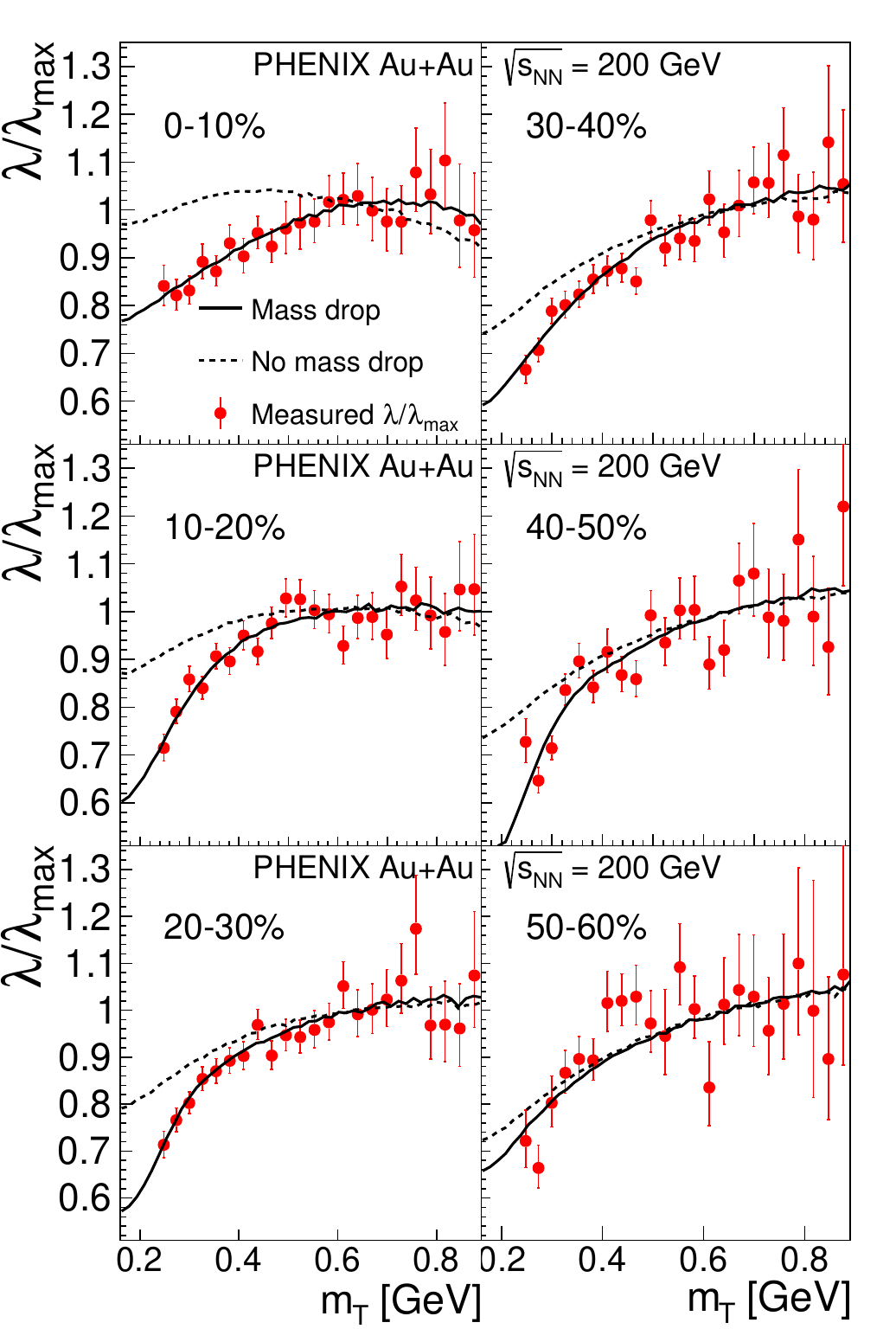}
    \caption{Comparison of the data to the optimal fits of the six 
centrality classes (solid line) and the case where the mass of the 
$\etaprime$ meson in hot hadronic matter is equal to the mass of the 
$\etaprime$ meson in vacuum (dashed line).}
    \label{fig:best-fits}
\end{figure}

Our results suggest that a significant, centrality independent, 
in-medium mass drop of the $\etaprime$ is not inconsistent with the 
present measurements. In Figs.~\ref{fig:final_B} 
and~\ref{fig:final_meta} the optimal values of $B^{-1}_{\etaprime}$ and 
$m^*_{\etaprime}$, respectively, are shown as function of $\Npart$. The 
average value of the modified $\etaprime$ mass is 
$m^*_{\etaprime}=581^{+12}_{-20}{\rm (stat)}^{+205}_{-91}{\rm (syst)}$ 
MeV. In the case of the inverse-slope parameter, the point corresponding 
to the 0\%--10\% centrality bin was omitted from the constant fit, i.e, 
the average value indicated in Fig.~\ref{fig:final_B}, represents the 
average value of the five remaining points, that is 
$B_{\etaprime}^{-1}=56^{+22}_{-14}{\rm (stat)}^{+190}_{-31}{\rm (syst)}$ 
MeV. In Fig.~\ref{fig:final_B} the solid [black] line denotes this 
centrality-averaged value. In at least five centrality classes, from 
0\%--10\% to 40\%--50\%, the measured $\lambda(\mT)/\lambda_{\rm max}$ 
functions are found to be consistently described by considering the 
suppressed mass of the $\etaprime$. However, in the most-peripheral 
centrality class (50\%--60\%) an unmodified $\etaprime$ mass cannot be 
excluded. In Fig.~\ref{fig:final_meta}, the dashed [red] line indicates the 
centrality-averaged value of the in-medium modified mass of the 
$\etaprime$ meson.  This observation of minimal mass modification can 
be interpreted as being due to the lack of enough hot and dense matter. The 
data are not inconsistent with Monte-Carlo simulations with vanishing 
differences between $m^*_{\etaprime}$ and $m_\eta{\approx}548$ MeV.

\begin{figure}
\includegraphics[width=1.0\linewidth]{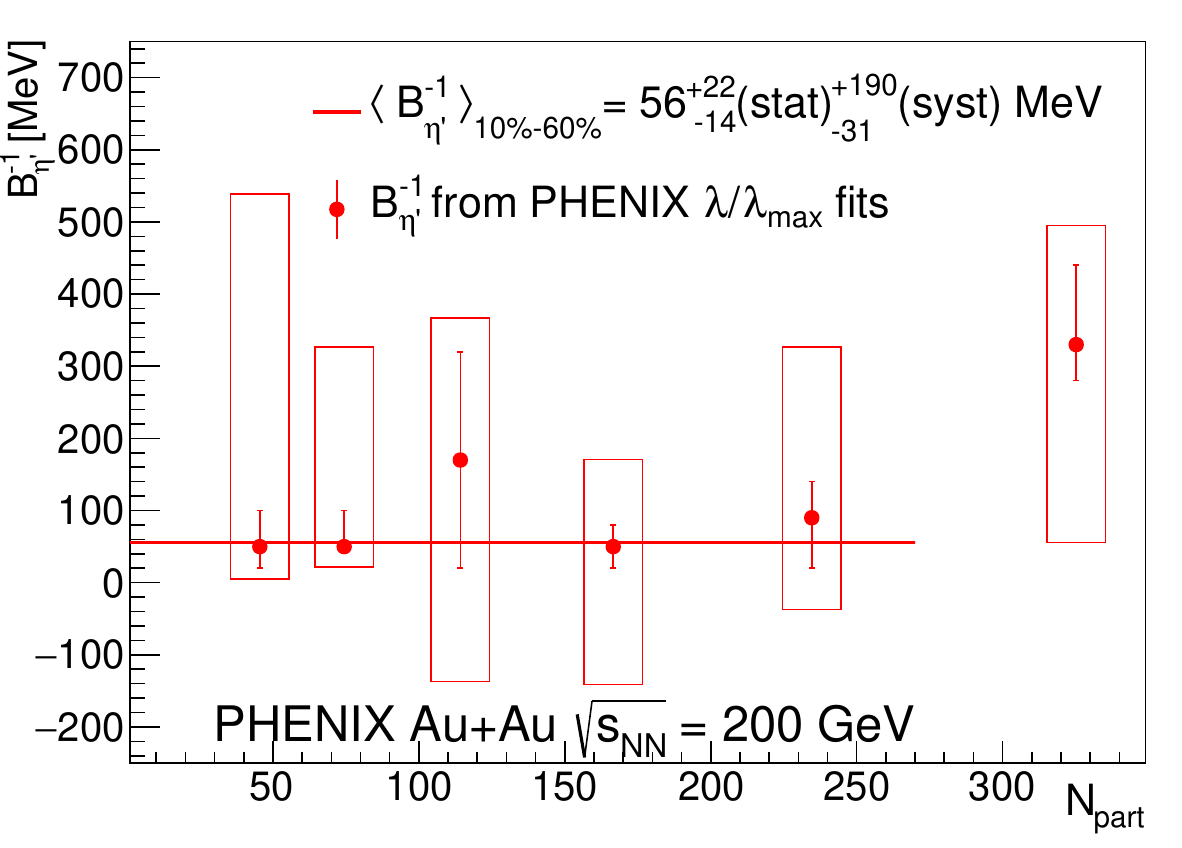}
\caption{The best values of $B_{\etaprime}^{-1}$ spectrum 
of the $\etaprime$ condensate in the six centrality classes.}
\label{fig:final_B}
\end{figure}

\begin{figure}
\includegraphics[width=1.0\linewidth]{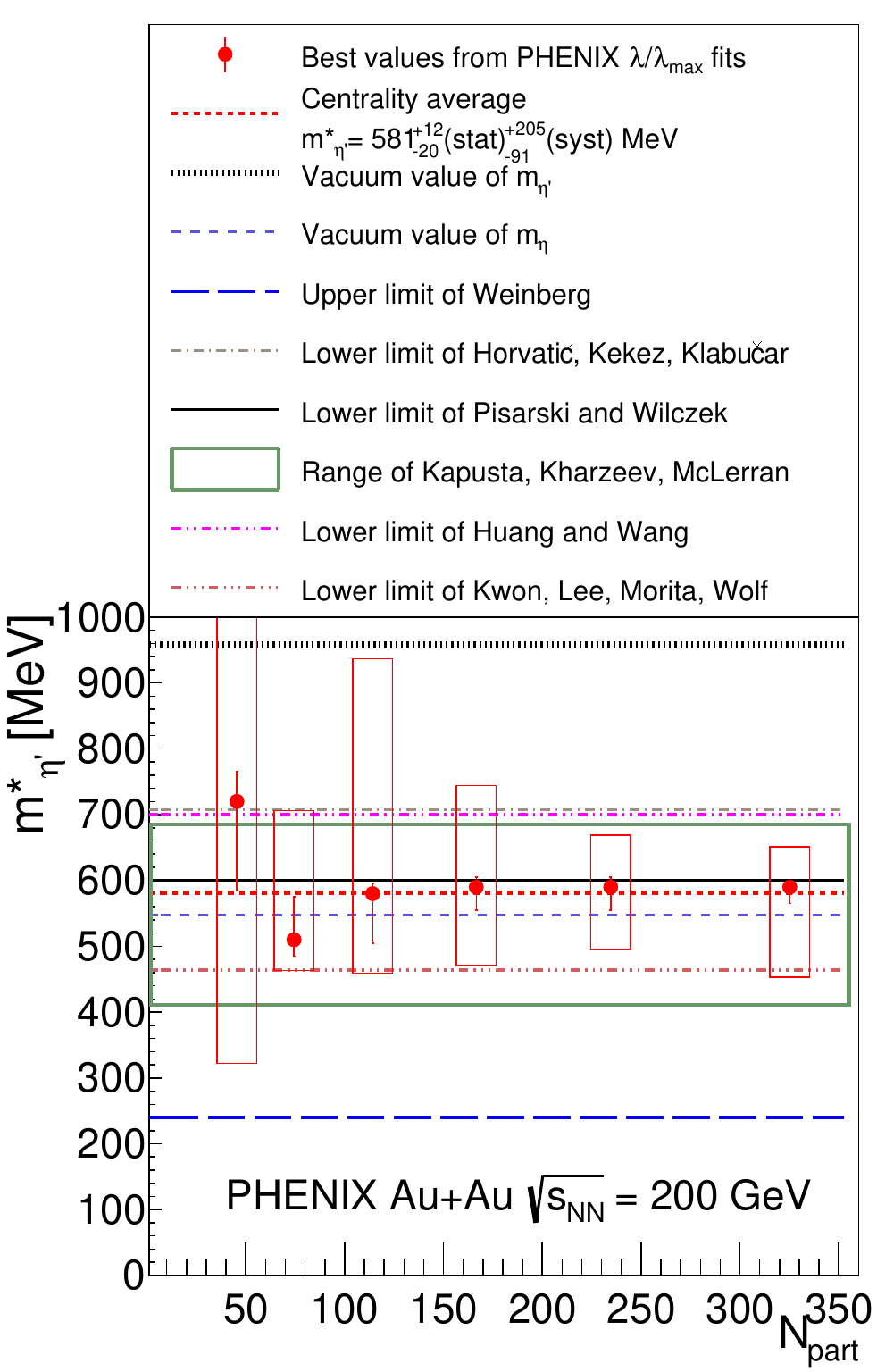}
\caption{The centrality dependence of the best values of the in-medium 
mass of the $m^*_{\eta^{\prime}}$ are shown with full circle symbols, 
together with statistical (vertical lines) and systematic (red boxes) 
uncertainties.  The average (using only statistical uncertainties) of 
the values of $m^*_{\etaprime}$ found in the six centrality intervals is 
indicated by a dashed (red) line. The fitted values are compared to 
theoretical predictions of Weinberg~\cite{Weinberg:1975ui}, Horvati\'c, 
Kekez and Klabu\v{c}ar ~\cite{Horvatic:2018ztu}, Pisarski and 
Wilczek~\cite{Pisarski:1983ms}, Kapusta, Kharzeev and 
McLerran~\cite{Kapusta:1995ww}, Huang and Wang~\cite{Huang:1995fc}, and 
Kwon, Lee, Morita, and Wolf~\cite{Kwon:2012vb}.}
\label{fig:final_meta}
\end{figure}

These are the first, centrality-dependent experimental results that 
suggest the suppression of the $\etaprime$ meson mass in a hot and dense 
medium. Figure~\ref{fig:final_meta} compares these results to some of 
the well-known theoretical predictions for the modified $\etaprime$ 
mass:

\begin{itemize}

\item
The Weinberg upper limit, $m^*_{\etaprime} \leq \sqrt{3} m_{\pi}$, 
suggested in Ref.~\cite{Weinberg:1975ui} clearly overestimates 
significantly, in each centrality bin, the possible in-medium mass drop 
of the $\etaprime$ particles.

\item
Recent calculations by Horvati\'c, Kekez, and Klabu\v{c}ar 
(HKK)~\cite{Horvatic:2018ztu}, based on the calculations of 
Witten-Veneziano equation and the generalization by Shore, evaluated the 
properties of the $\eta$ and $\etaprime$ mesons at high temperatures, 
when the $U_A(1)$ and the chiral symmetry breaking are considered 
together. A substantial decrease in the $\etaprime$ mass around the 
chiral transition temperature was obtained in 
Ref.~\cite{Horvatic:2018ztu}, but there was no decrease in the $\eta$ 
mass. The new HKK results are an improvement on the earlier results of 
Ref.~\cite{Csorgo:2009pa,Vertesi:2009wf}. The lower limits, which are 
shown by the [light-green] dashed-dotted line in Fig.~\ref{fig:final_meta}, 
lie above our results except in the 50\%--60\% centrality class, where 
the uncertainty on our value of $m^*_{\etaprime}$ is particularly large. 
This is also true of the similar limit of Huang and 
Wang~\cite{Huang:1995fc}. These models are in a modest tension with our 
results.

\item 
The Pisarski-Wilczek lower limit of 600 MeV, as determined from Fig.~1 
of Ref.~\cite{Pisarski:1983ms}, is shown as a solid [black] line. It is, 
within uncertainty, consistent, with the fitted values.

\item
The significant in-medium mass drop of the $\etaprime$ meson was related 
to the restoration of $U_A(1)$ symmetry and described as the return of a 
Goldstone-boson by Kapusta, Kharzeev and McLerran 
(KKM)~Ref.~\cite{Kapusta:1995ww}. They have given a broad range for the 
possible in-medium mass of the $\etaprime$ meson in case of a {\it 
partial} ${\rm U}_A(1)$ symmetry restoration, namely $411$ MeV $\leq 
m^*_{\eta^{\prime}} \leq 685$ MeV shown as a [green] box in 
Fig.~\ref{fig:final_meta}. Our results lie within this range.

\item
Kwon, Lee, Morita, and Wolf (KLMW) also utilized a 
temperature-dependent, generalized Witten-Veneziano relation to obtain 
a nearly 50\% decrease in the mass of the $\etaprime$ meson in hot and 
dense hadronic medium, as a consequence of the restoration of $U_A(1)$ 
symmetry~\cite{Kwon:2012vb}. Their 2012 prediction of $m^*_{\etaprime} 
\geq 464$ MeV, shown as a triple dotted-dashed line in 
Fig.~\ref{fig:final_meta}, is consistent with our results in each 
investigated centrality class.

\end{itemize}



\section{Summary and Conclusion}
\label{sec:conclusions} 

This paper presents measurements of the two-pion BEC function in Au$+$Au 
collisions at \sqsntwo = 200 GeV using data recorded in 2010 by the 
PHENIX experiment. L{\'e}vy-stable distributions are utilized to 
characterize the data and determine the transverse-mass and centrality 
dependence of the L{\'e}vy parameters.

The L{\'e}vy parameterization is found to give a statistically 
acceptable description of the data with the L\'evy exponent, $\alpha$, 
significantly greater than 1 and less than 2. This exponent is 
determined in 23 $\mT$ and in 6 centrality bins and is observed to 
be well described with its $\mT$-averaged value in every centrality bin. 
However, the transverse-mass-averaged values do depend on centrality.

The L{\'e}vy-scale parameter $R$ is proportional to the HWHM of the 
source distribution, with a coefficient of proportionality that depends 
on the L\'evy exponent $\alpha$. The behavior of its inverse square as 
function of $\mT$ that is observed can be described by an affine linear 
fit, whose parameters are examined as a function of centrality. The 
parameter $A$, which can be related to the transverse velocity of the 
expansion, shows a trend similar to hydrodynamical 
predictions~\cite{Csorgo:1995bi,Chapman:1994ax}. However, the assumption 
of local thermalization results in the $\alpha=2$ special case (see, 
e.g., Refs.~\cite{Csorgo:1995bi,Akkelin:1995gh,Chapman:1994ax}), which 
is in significant contrast to our observations of $\alpha<2$ in each of 
the investigated centrality and transverse-mass bins.

Within statistical uncertainties, the parameter $B$, is consistent with 
zero or with a slightly negative value. A negative value of $B$ is 
possible if the Cooper-Frye freeze-out terms are also taken into 
account~\cite{Csorgo:1999sj}. This result may indicate a source that 
includes local thermalization with hydrodynamical expansion, followed by 
rescattering and decays of resonances. However, such nonequilibrium, 
scale-dependent features typically would result in deviations from the 
L\'evy shape and from the applicability of generalized central-limit 
theorems. But first-order deviations from the L\'evy-stable source 
distributions using the expansion method of 
Refs.~\cite{Csorgo:2000pf,Novak:2016cyc,Csorgo:2018uyp} were found to be 
consistent with zero. It is theoretically challenging to explain 
simultaneously the measured value of $\alpha$, which is found to be 
significantly less than the Gaussian value of 2, and the $\mT$ 
dependence of the L\'evy scale parameter $R$, which follows a 
hydrodynamically predicted scaling (see, e.g., 
Ref.~\cite{Kincses:2022eqq}).

The connection to the initial geometry is supported by the linearity of 
the parameter $R$ as a function of $\Npart^{1/3}$ in any given $\mT$ 
bin. The precise characterization of the correlation functions and the 
prudent handling of the Coulomb final-state interaction make it possible 
to determine in detail the $\mT$ and centrality dependence of the 
$\lambda(\mT)$ intercept parameter, as well as of its normalized form, 
$\lambda(\mT)/\lambda_{\rm max}$. In the trends of the latter it can be 
qualitatively observed that there is a low-$\mT$ suppression of 
$\lambda(\mT)/\lambda_{\rm max}$ in every centrality bin and that the 
characteristics are more or less the same, i.e., the suppression is 
consistent with the hypothesis of centrality independence.

To quantify the suppression pattern, the Gaussian width and amplitude 
parameters, $\sigma$ and $H$, were introduced. These parameters are 
observed to be centrality independent, except in the 50\%--60\% case, 
where the uncertainty on $H$ increases and does not allow a 
statistically significant conclusion. This independence is helpful to 
rule out or validate models with predictions about this behavior. The 
pion laser model, for example, predicts a strong centrality and 
multiplicity 
dependence~\cite{Zimanyi:1997ur,Csorgo:1997us,Csorgo:2009pa}.

In Ref.~\cite{Vance:1998wd} it was observed that the amount of 
suppression of $\lambda(\mT)$ and hence of $\lambda(\mT)/\lambda_{\rm 
max}$ at low $\mT$ can be an experimentally observable signal of the 
(partial) restoration of $U_A(1)$ symmetry and a measure of in-medium 
reduction of the mass of the anomalously heavy $\etaprime$ mass. An 
approximately constant trend of the intercept parameter $\lambda$ as a 
function of $\mT$ was observed in small systems at lower energies, 
($E_{\rm LAB} = 150$ AGeV~\cite{NA61SHINE:2023qzr,NA44:1994dmh}), which 
is consistent with vanishing suppression ($H=0$). These results are both 
qualitatively and quantitatively different from the presence of the 
suppression ($H_0 = 0.42 \pm 0.02$(stat) ), which is within 
uncertainties independent of the centrality in $\sqrt{s_{NN}}=200$ GeV 
Au$+$Au collisions at RHIC. This entirely experimental data-based, 
Monte-Carlo-independent observation suggests that the signal 
characterized by the value of $H$ is dependent on the energy and/or the 
system size. Further detailed measurements are needed to determine the 
energy and system size, where $H$ becomes nonvanishing for the first 
time, starting from zero in the small Be+Be collisions at the relatively 
low energy of $E_{\rm LAB}=150$ AGeV.

There could also be alternative or competing effects that could modify 
the correlation strength, see e.g. Ref.~\cite{Csanad:2020qtu}.  Further 
theoretical studies are needed to explain these data using other 
methods. However, one of the possible explanations relates the 
observations to restoration of $U_A(1)$ symmetry in a hot and dense 
hadronic matter as detailed in 
Refs.~\cite{Vance:1998wd,Csorgo:2009pa,Vertesi:2009wf}.

It is clear that our observed suppression of the 
$\lambda(\mT)/\lambda_{\rm max}$ parameter at lower transverse mass is 
not inconsistent with the in-medium mass modification of the $\etaprime$ 
mesons, related to $U_A(1)$ symmetry restoration in hot hadronic matter. 
This relation was cross-checked with the help of detailed Monte-Carlo 
simulations, and creating $\chi^2$ and CL maps that compared simulations 
which allowed in-medium $\etaprime$ mass modification in hot and dense 
hadronic matter. As detailed in Section~\ref{sec:simulations}, and also 
in Table~\ref{tab:results_m_B} of the Appendix it is shown that for each 
of the considered centrality classes the best value of the in-medium 
mass of the $\etaprime$ meson, $m_{\eta'}^* =581^{+12}_{-20}{\rm 
(stat)}^{+205}_{-91}{\rm (syst)}$ MeV.

This mass is, within the uncertainties of this indirect measurement, the 
same as the Particle Data Group (PDG) value of the $\eta$ meson, $m_\eta 
= 547.86 \pm 0.02$ MeV~\cite{ParticleDataGroup:2022pth}. This 
observation suggests that the return of the so-called prodigal Goldstone 
boson~\cite{Kapusta:1995ww} and the restoration of the $U_A(1)$ symmetry 
is not inconsistent with our measurements.

However, our measurements are inconsistent with Monte-Carlo simulations 
that utilize the PDG value of the $\etaprime$ mass, $m_{\etaprime} = 
957.78 \pm 0.06$ MeV, which does not allow for mass modification, except 
in the most peripheral, (50\%--60\%) centrality class. Several 
theoretical predictions from 
Refs.~\cite{Weinberg:1975ui,Pisarski:1983ms,Kapusta:1995ww,Kwon:2012vb,Horvatic:2018ztu,Huang:1995fc} 
are compared to the results of our $\chi^2 $ maps. These comparisons can 
be summarized as follows:

\begin{itemize}

\item
   The Kapusta-Kharzeev-McLerran prediction~\cite{Kapusta:1995ww} is in 
agreement with our measurements in each investigated centrality class.

\item
   The lower limit of Kwon, Lee, Morita, and Wolf~\cite{Kwon:2012vb} is 
also consistent with our measurement in each investigated centrality 
class.

\item 
   Our measured centrality-average value of $m^*_{\etaprime}$ is 
slightly below, but consistent with, the lower limit predicted by 
Pisarski and Wilczek~\cite{Pisarski:1983ms}.

\item
   However, the upper limit of Weinberg~\cite{Weinberg:1975ui} is 
several standard deviations below the central values obtained in each 
investigated centrality class.

\item
    The lower limit predictions of Horvati\'c, Kekez and 
Klabu\v{c}ar~\cite{Horvatic:2018ztu} and of Huang and 
Wang~\cite{Huang:1995fc} are excluded except in the 50\%--60\% 
centrality class.

\item
    Our results also suggest that the prediction of 
Ref.~\cite{Kovacs:2021kas} slightly underestimates the in-medium mass 
change of the $\etaprime$.

\end{itemize}

However, the lack of in-medium $\etaprime$ mass modification is not 
consistent with our measurements, except in the 50\%--60\% centrality 
class, as discussed and detailed in the Appendix. Thus, these PHENIX 
results provide the most detailed, centrality-dependent constraints for 
future theoretical studies on $U_A(1)$ symmetry restoration in hot and 
dense hadronic matter. These results also exhibit an unprecedented 
selection power by excluding certain models in certain centrality 
classes. Thus these indirect PHENIX measurements provide important 
constraints and insights to future studies of (partial) $U_A(1)$ 
symmetry restoration in hot and dense hadronic matter.

For future experimental studies, these results emphasize the need for 
direct measurements of identified $\etaprime$ spectra in high-energy 
heavy-ion collisions at RHIC and at the Large Hadron Collider, in 
particular in the very soft, $\pT\ \leq 300$ MeV kinematic domain at 
midrapidity. Direct experimental observations of enhanced production of 
soft $\etaprime$ mesons seem to be particularly difficult, due to the 
expected backgrounds. Huge backgrounds, e.g. in the $\etaprime 
\rightarrow \gamma\gamma$ decay channel, are expected, e.g., from $\pi^0 
\rightarrow \gamma\gamma $ decays. Thus a direct observation of 
in-medium $\etaprime$ mass modification seems to be experimentally 
challenging, but based on the indirect results summarized here, are also 
expected to be particularly rewarding.

\acknowledgments

We thank the staff of the Collider-Accelerator and Physics
Departments at Brookhaven National Laboratory and the staff of
the other PHENIX participating institutions for their vital
contributions.
We acknowledge clarifying and inspiring discussions with S\'andor 
Hegyi, Dubravko Klabu{\v c}ar, Robert Pisarski and Gy{\"o}rgy Wolf 
concerning the interpretation of their published results.  
S.L. thanks IFJ PAN, Krakow, Poland for their encouragement to 
finish this work.
We acknowledge support from the Office of Nuclear Physics in the
Office of Science of the Department of Energy,
the National Science Foundation,
Abilene Christian University Research Council,
Research Foundation of SUNY, and
Dean of the College of Arts and Sciences, Vanderbilt University
(U.S.A),
Ministry of Education, Culture, Sports, Science, and Technology
and the Japan Society for the Promotion of Science (Japan),
Conselho Nacional de Desenvolvimento Cient\'{\i}fico e
Tecnol{\'o}gico and Funda\c c{\~a}o de Amparo {\`a} Pesquisa do
Estado de S{\~a}o Paulo (Brazil),
Natural Science Foundation of China (People's Republic of China),
Croatian Science Foundation and
Ministry of Science and Education (Croatia),
Ministry of Education, Youth and Sports (Czech Republic),
Centre National de la Recherche Scientifique, Commissariat
{\`a} l'{\'E}nergie Atomique, and Institut National de Physique
Nucl{\'e}aire et de Physique des Particules (France),
J. Bolyai Research Scholarship, EFOP, HUN-REN ATOMKI, NKFIH, MATE KKP
and OTKA (Hungary),
Department of Atomic Energy and Department of Science and Technology (India),
Israel Science Foundation (Israel),
Basic Science Research and SRC(CENuM) Programs through NRF
funded by the Ministry of Education and the Ministry of
Science and ICT (Korea).
Ministry of Education and Science, Russian Academy of Sciences,
Federal Agency of Atomic Energy (Russia),
VR and Wallenberg Foundation (Sweden),
University of Zambia, the Government of the Republic of Zambia (Zambia),
the U.S. Civilian Research and Development Foundation for the
Independent States of the Former Soviet Union,
the Hungarian American Enterprise Scholarship Fund,
the US-Hungarian Fulbright Foundation,
and the US-Israel Binational Science Foundation.


\section*{Appendix: Details of the Monte-Carlo simulations}
\label{sec:appendix}

Monte-Carlo simulations with scenarios that allow for (or exclude) a 
possible in-medium mass modification of the $\etaprime$ meson and their 
comparisons to data were performed in 
Refs.~\cite{Vance:1998wd,Csorgo:2009pa,Vertesi:2009wf} and in the 
previous PHENIX paper (Ref.~\cite{PHENIX:2017ino}) by utilizing standard 
$\chi^2 $ and CL maps. Similar, but now centrality-dependent simulations 
are summarized in this Appendix. These comparisons result in $\chi^2$ 
and CL maps that are utilized to determine the $\chi^2$ minimum (or CL 
maximum) yielding the best values, as well as the statistical and the 
systematic uncertainties of a possible in-medium mass-drop of the 
$\etaprime$ meson. Figure~\ref{fig:CL-maps} shows an example of such a 
comparison, where each bin corresponds to a comparison of Monte-Carlo 
simulations of resonance decay chains with our data on 
$\lambda(\mT)/\lambda_{{\rm max}}$ as a function of centrality, as 
detailed in Section~\ref{sec:simulations}.
 
Each panel of Fig.~\ref{fig:CL-maps} contains the optimal values of 
$m^*_{\etaprime}$ and $B_{\etaprime}^{-1}$ in the scanned region. For 
our simulations, a unique $\chi^2$ minimum is found, in contrast to 
earlier studies in Refs.~\cite{Csorgo:2009pa,Vertesi:2009wf}. By now, it 
is well known, that the dominant mechanism for soft-particle production 
is a thermal one in \sqsntwo = 200 GeV Au+Au collisions.

Earlier, a significantly broader class of models was considered in 
Refs.~\cite{Csorgo:2009pa,Vertesi:2009wf} that resulted in two, 
characteristically different minima in the 0\%--30\% centrality class of 
the same reaction. Accordingly, Refs.~\cite{Csorgo:2009pa,Vertesi:2009wf} 
evaluated the exclusion limits, focusing in particular on the smallest 
possible in-medium mass drop of the $\etaprime$ meson, which is required 
to describe the data with at least 0.1\%~CL in any of the considered 
Monte-Carlo simulations.

Our systematic investigations include an estimation of the exclusion 
limits for the lowest- and largest-possible values of the in-medium mass 
of the $\eta'$ meson.  The systematic variations are consistent, i.e. 
all the systematic variations result in essentially the same minimum. 
Thus, in contrast to Refs.~\cite{Csorgo:2009pa,Vertesi:2009wf}, our 
indirect observation of in-medium $\eta'$ mass modification is used with 
the best value and its statistical and systematic uncertainties of 
$m^*_{\etaprime}$, rather than exclusion limits. 


Table~\ref{tab:results_m_B} shows the centrality-dependent, fitted 
values of the in-medium $\eta'$ mass $m^*_{\etaprime}$ and 
the inverse slope $B^{-1}_{\etaprime}$ with statistical and 
systematic uncertainties.
The exclusion limits of these parameters were evaluated, but 
they do not result in statistically significant new ranges.

\begin{figure}
    \includegraphics[width=1.0\linewidth]{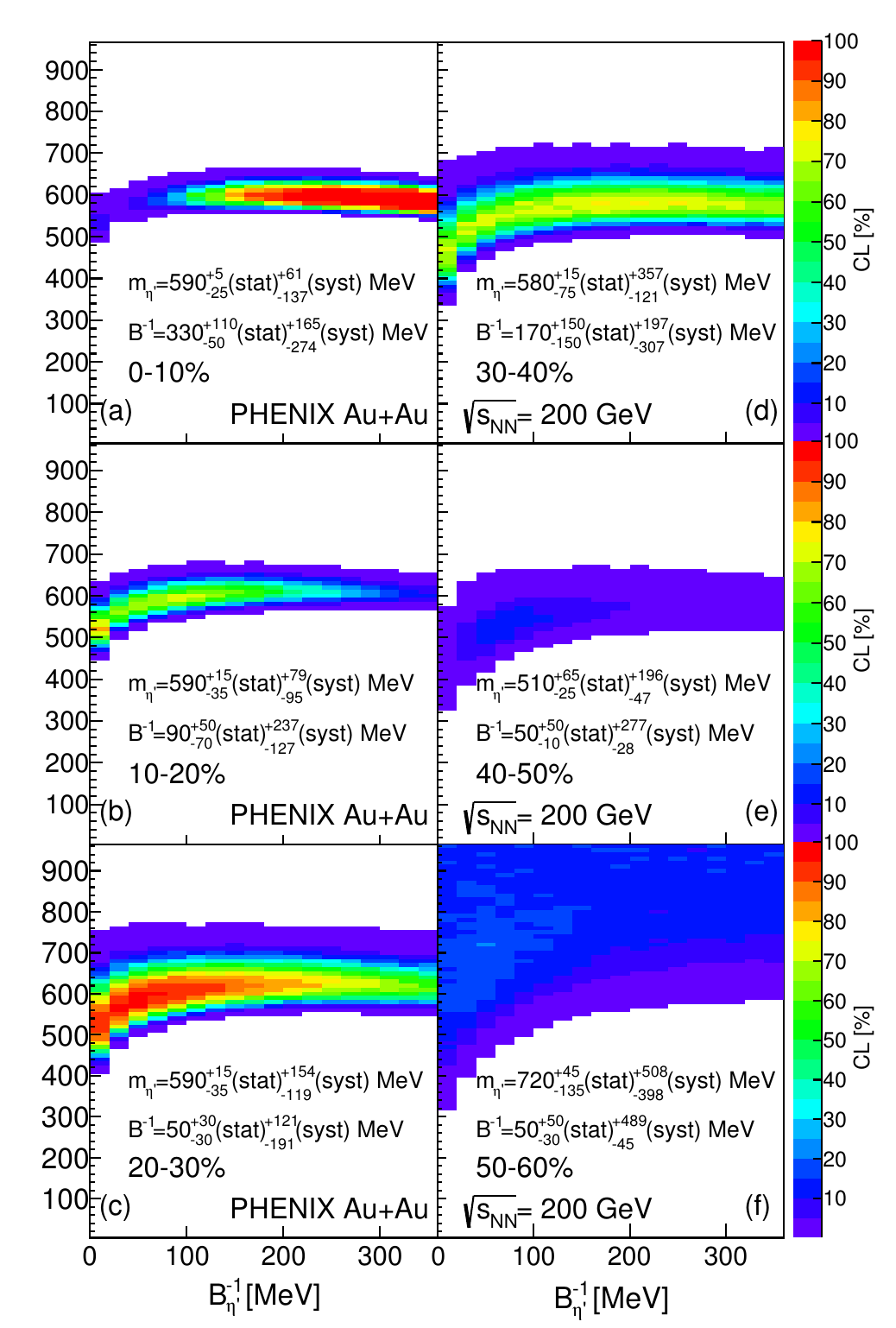}
    \caption{The CL maps of the $m^*_{\etaprime}$ and 
$B_{\etaprime}^{-1}$ scans in the six centrality classes.}
    \label{fig:CL-maps}
\end{figure}

The $\etaprime$ spectrum has six parameters, which are considered as 
inputs to the $\lambda(\mT)$ simulations. The two most important inputs 
are the in-medium mass of the $\etaprime$, $m^*_{\etaprime}$, and the 
characteristic-slope parameter, $B_{\etaprime}^{-1}$, of the $\etaprime$ 
mesons that are emitted with PDG mass after the decay of the condensate. 
The second parameter characterizes a second component, a low-$\pT$ part 
of the $\etaprime$ spectrum and determines the slope of the suppression 
of the $\lambda$ parameter at 
low-$\pT$~\cite{Vance:1998wd,Csorgo:2009pa,Vertesi:2009wf}. These two 
parameters determine the shape of the observed dip in the low-$\mT$ 
region of the $\lambda(\mT)$ function: a strongly reduced $\etaprime$ 
mass causes a dip in the low-$\mT$ part of the $\lambda(\mT)$ function, 
while $B_{\etaprime}^{-1}$ controls the slope of this dip, see 
Refs.~\cite{Csorgo:2009pa,Vertesi:2009wf}. These two parameters are 
considered as fit variables. The values of four nuisance parameters 
$\alpha_{\rm th}$, $\Tf$, $\langle u_{T}\rangle$, and $T_{\rm cond}$, 
are treated as constants in the fit, but are varied to evaluate the 
systematic uncertainties of the fitted values.

The $\chi^2$ scans provide fine grids of CL in each centrality class. 
The in-medium mass is scanned with 10 MeV steps between 0 and 958 MeV 
(the vacuum mass of $\etaprime$) and $B_{\etaprime}^{-1}$ with 20 MeV 
steps between 0 and 360 MeV. The minimum value of $\chi^2$ (or maximum 
value of CL) is determined, which provides the most likely values of 
$m^*_{\etaprime}$ and $B_{\etaprime}^{-1}$ and their statistical 
uncertainties, as well as the CL of the fits to the data, which are 
compared to the data in Fig.~\ref{fig:best-fits}. The values of the 
four ``constant'' parameters, $\alpha_{{\rm th}}$, $\Tf$, 
$\langle\uT\rangle$, and $T_{\rm cond}$, are varied to evaluate the 
systematic uncertainties of the fitted values:

\begin{description}

\item[$\alpha_{{\rm th}}$]
The centrality-dependent invariant single particle spectra of positively 
and negatively charged kaons as well as protons and antiprotons of 
Ref.~\cite{PHENIX:2003iij} are fitted with the formula
\begin{align}
N(\mT) = C {\mT}^{\alpha_{{\rm th}}} \exp\left(- \frac{\mT}{\Teff}\right)
\end{align}
where $C$ is a normalization constant. The polynomial exponent of the 
thermal spectrum is denoted by $\alpha_{{\rm th}}$, to distinguish it 
from the L\'evy-exponent $\alpha$. Note that the parameter $\alpha_{{\rm 
th}}$ of this paper is the same, as the $\alpha$ parameter of 
Refs.~\cite{Vance:1998wd,Csorgo:2009pa,Vertesi:2009wf}. The exponent 
$\alpha_{{\rm th}} = 1 - d/2$, where $d$ is the number of spatial 
dimensions of the expansion; hence its allowed range is $1 \leq d \leq 
3$. The value of $\alpha_{th}$ is fixed to $\alpha_{{\rm th}}=0$, which 
corresponds to an effectively two-dimensional 
expansion~\cite{Csorgo:1994in,Csorgo:1995bi}. The same value was used in 
Ref.~\cite{PHENIX:2003iij} by the PHENIX experiment when obtaining 
$\Teff$. Its value is assumed to be independent of centrality and is 
varied in the systematic studies between $1/2$ and $-1/2$. Also similar 
to Ref.~\cite{PHENIX:2003iij}, good quality exponential fits are 
obtained, with CL $\geq 0.1$\%, in each centrality class for charged 
kaons, protons, and antiprotons in the transverse-mass range of 
$0.1\leq\mT-m\leq{1.0}$ GeV.  Similar to Ref.~\cite{PHENIX:2003iij}, the 
mass dependence of the slopes is also well described with affine linear 
fits: $\Teff(m) =\Tf+m \langle \uT \rangle^2$, but here these fits have 
a good CL with CL $\geq 0.1$\% in each centrality class.

\item[$\Tf$]
The kinetic freeze-out temperature is denoted by $\Tf$. Significant 
centrality dependent results for the kinetic freeze-out temperature are 
obtained here. In particular, the value of $\Tf$ is significantly lower 
for the 0\%--10\% centrality class than for other centrality classes. 
For peripheral centrality classes, the value of $\Tf$ increases, 
reaching its upper limit, the value of the chemical freeze-out 
temperature $T_{{\rm chem}}$. Our results for the variations of $T_f$ 
and $\langle u_T\rangle$ with centrality are shown as inserted values in 
Fig.~\ref{fig:lambdascaling}. Fixing $\Tf$ to a centrality-independent 
constant, results in the above affine linear-mass-dependent coefficient 
fits, similar to Ref.~\cite{PHENIX:2003iij}, having CLs that are too 
small, well below the $0.1$\% $\leq {{\rm CL}}$ threshold values. A 
possible reason for a lower kinetic freeze-out temperatures $\Tf$ in the 
0\%--10\% centrality class is that a larger volume may cool remarkably 
longer and to a lower temperature, as noted by Hama and 
Navarra~\cite{Hama:1990tx}.

\item[$\langle\uT\rangle$] The average radial flow is denoted by 
$\langle\uT\rangle$. This parameter influences the overall slope of the 
$\lambda(\mT)$ distribution at higher transverse mass and results in a 
centrality dependent expectation for 
$\lambda(\mT)/\lambda_\textmd{max}$. Note that the pion halo contains 
the decay products of long-lived resonances, which include the decay 
products of $\omega$, $\eta$, $\etaprime$ and $K_s^0$ mesons. It is thus 
important to simulate these decay chains precisely and in agreement with 
available experimental data. The best values of $\Tf$ and $\langle 
\uT\rangle$ are obtained by simultaneous fits to the slope parameters of 
the positively and negatively charged kaons, protons, and antiprotons 
with the formula $\Teff = \Tf + m \langle \uT\rangle^2$ where $m$ is the 
mass of the meson or baryon.  These parameters are systematically varied 
within the uncertainties allowed by these fits, shifting and fixing one 
of the $(\Tf,\langle \uT \rangle)$ pair by one standard deviation and 
refitting the other parameter to take into account their covariation.  
The invariant spectra of $K^{\pm}$ and those of protons and antiprotons 
are simulated using fits with good CL.  The mass scaling of these 
spectra is then utilized.  Nearly the same are the mass of the $\eta$ 
meson compared to charged kaons and the mass of the $\etaprime$ meson 
compared to protons and antiprotons.  This gives a good basis to 
describe well the null effect, which is the no in-medium mass 
modification scenario of the $\eta$ and $\etaprime$ spectrum. This 
method has been tested and the test was published in Fig. 11 of 
Ref.~\cite{Vertesi:2009wf}. Extrapolating the mass-scaled simulations to 
the PHENIX acceptance successfully reproduces the $\eta$ spectrum 
measured in $\sqrt{s_{_{NN}}}=200$ GeV Au$+$Au 
collisions~\cite{PHENIX:2006avo} in the $\mT - m_\eta \geq 1.75$ GeV 
range.

\item[$T_{\rm cond}$] The effective temperature of the in-medium 
$\etaprime$ condensate is denoted by $T_{{\rm cond}}$, as in 
Refs.~\cite{Vance:1998wd,Csorgo:2009pa,Vertesi:2009wf}.  This parameter, 
together with the in-medium mass of the $\etaprime$ mesons, controls the 
thermal enhancement of the $\etaprime$ mesons in the medium. The smaller 
this number, the smaller the $\etaprime$ mass drop for a given 
$\lambda(\mT)/\lambda_{\rm max}$ measurement.  Hence, the most 
conservative assumption is made for the default value, $T_{\rm cond} = 
\Tf$. As this temperature is inside the hot and dense, but already 
hadronic matter, part of our systematic studies is to vary $T_{\rm 
cond}$ within the allowed range of $T_{\rm chem} \geq T_{\rm cond} \geq 
T_{\rm f}$. As the conservative default value lies at one of the edges 
of the allowed interval, this choice results in a more than usually 
asymmetric uncertainty distribution on the physical fit parameters, such 
as the in-medium mass of the $\etaprime$ and the slope parameter of the 
$\etaprime$ after the decay of the condensate, $m^*_{\etaprime}$ and 
$B_{\etaprime}^{-1}$. From these systematic uncertainty studies, the 
in-medium $\etaprime$ mass is frequently observed to be smaller than our 
quoted best values, but cannot easily be larger than the best values 
because its upper uncertainties are much smaller than its lower 
uncertainties.

\end{description}

\begin{figure}[tbh]
    \includegraphics[width=1.0\linewidth]{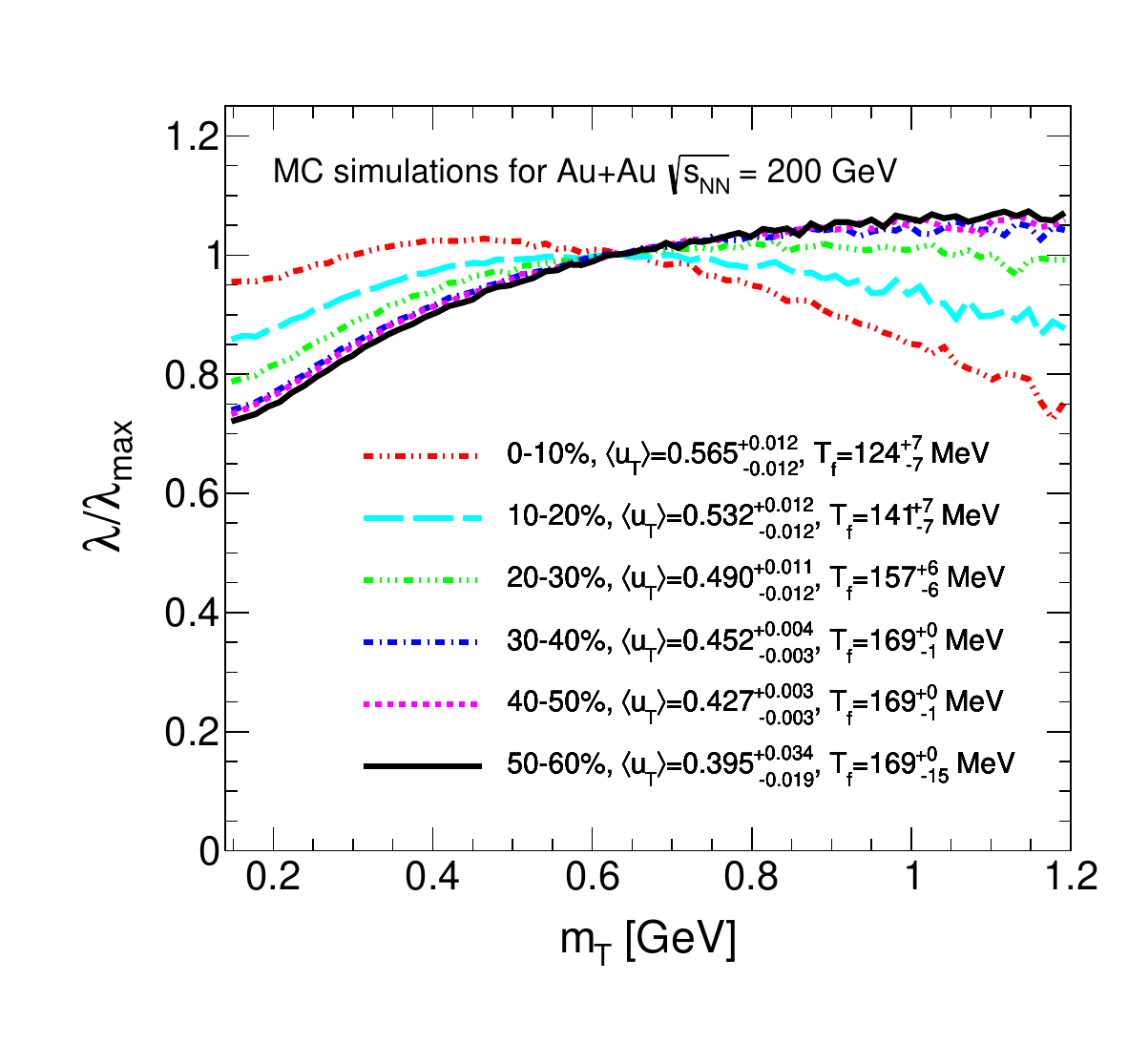}
    \caption{Centrality dependent expectations for 
$\lambda(\mT)/\lambda_{{\rm max}}$, based on Monte-Carlo simulations 
without any in-medium $\etaprime$ modification.
    }
    \label{fig:lambdascaling}
\end{figure}

For each of the parameters $\alpha_{{\rm th}}$, $\Tf$, $\langle 
\uT\rangle$, and $T_{{\rm cond}}$, the $\chi^2$ scans and the 
corresponding CL maps are redone for the variations. The relative 
difference between the default and the alternative setting is 
calculated. The final systematic uncertainty is the quadratic sum of 
these differences.

Additionally, several other cases are also investigated, including two 
special cases, $\Tf = 140$ MeV fixed, independent of centrality, and 
$\Tf = 177$ MeV fixed, also independent of centrality. The first case 
corresponds to Landau's calculation of the freeze-out temperature (equal 
to the pion mass, mass of the lightest neutral quanta). The second case 
is consistent with the PHENIX publication on charged pion, kaon, and 
(anti)proton spectra~\cite{PHENIX:2003iij}. The latter choice is found 
to be inconsistent with $T_{{\rm cond}} \geq \Tf$. The former gives 
results that are within the quoted systematic uncertainties.

Similarly to Ref.~\cite{PHENIX:2017ino}, the statistical uncertainty of 
$\lambda_{\rm max}$ is treated as a normalization uncertainty. Both this 
uncertainty ($\approx{1}$\%) and the systematic uncertainty 
($\approx{1}$\%) caused by the choice of $\mT$ range when calculating 
$\lambda_{\rm max}$ are negligible compared to other uncertainties, 
except in the 50\%--60\% centrality case of $B_{\etaprime}^{-1}$.  In 
that case, no significant in-medium mass modification is found.  Thus, 
$B_{\etaprime}^{-1}$ cancels from the results and cannot be precisely 
determined.  Therefore, these three negligibly small sources of 
systematic errors are not included here.

The simulation does not include the PHENIX detector system; hence, the 
experimental systematic uncertainties are accounted for by propagating 
the total systematic uncertainties of the measured 
$\lambda/\lambda_{{\rm max}}$. The values of the $m^*_{\etaprime}$ and 
$B_{\etaprime}^{-1}$ parameters with statistical and systematic 
uncertainties are shown in Fig.~\ref{fig:final_meta} and 
Fig.~\ref{fig:final_B}, respectively. The centrality-dependent results 
for the in-medium $\etaprime$ masses are shown on 
Fig.~\ref{fig:final_meta} and the corresponding numerical values are 
tabulated in Table~\ref{tab:results_m_B}.

\begin{table}[ht]
\caption{
\label{tab:results_m_B}
The centrality-dependent, fitted values, in [MeV], of the in-medium 
$\eta'$ mass $m^*_{\etaprime}$ and the inverse slope 
$B^{-1}_{\etaprime}$. The statistical (stat) and systematic (syst) 
uncertainties are also presented.
}
\begin{ruledtabular} \begin{tabular}{cccccccccc}
Centrality
   & & $m^*_{\eta'}$   & (stat) & (syst)
   & & $B^{-1}_{\eta'}$ & (stat) & (syst) & \\
\hline
0\%--10\%  & & 590 &  $^{+5}_{-25}$ & $^{+61}_{-137}$  &  & 330 & $^{+110}_{-50}$  & $^{+165}_{-274}$ & \\
10\%--20\% & & 590 & $^{+15}_{-35}$ & $^{+79}_{-95}$   &  & 90  & $^{+50}_{-70}$   & $^{+237}_{-127}$ & \\
20\%--30\% & & 590 & $^{+15}_{-35}$ & $^{+154}_{-119}$ &  & 50  & $^{+30}_{-30}$   & $^{+121}_{-191}$ & \\
30\%--40\% & & 580 & $^{+15}_{-75}$ & $^{+357}_{-121}$ &  & 170 & $^{+150}_{-150}$ & $^{+197}_{-307}$ & \\
40\%--50\% & & 510 & $^{+65}_{-25}$ & $^{+196}_{-47}$  &  & 50  & $^{+50}_{-10}$   & $^{+277}_{-28}$  & \\
50\%--60\% & & 720 & $^{+45}_{-135}$& $^{+508}_{-398}$ &  & 50  & $^{+50}_{-30}$   & $^{+489}_{-45}$  & \\
\end{tabular} \end{ruledtabular}
\end{table}

As mentioned in Section~\ref{sec:mt_dep}, the normalized intercept 
parameter dependence on $\mT$ and centrality in are investigated in 
detail. An explanation of the dependence is given in terms of radial 
flow and $\pT$ sharing among the pions arising in $\etaprime$ decay. The 
simulations suggest a $\lambda(\mT)/\lambda_{{\rm max}}$ curve, that 
changes monotonically with centrality. These changes appear related to 
the monotonic decrease of radial flow $\langle \uT\rangle$, coupled to a 
monotonic increase of the kinetic freeze-out temperature $\Tf$ as the 
collisions change from most central to more and more-peripheral 
collisions~\cite{Vance:1998wd}. The values of $\Tf$ and $\langle 
\uT\rangle$ are obtained from affine linear fits with $\Teff = \Tf + m 
\langle \uT\rangle^2$ to the slope parameters of the charged kaon, $K^+$ 
and $K^-$, as well as to the proton and the antiproton single-particle 
spectra. Particular attention is paid to the requirement that the single 
exponential fit to the single-particle spectra have acceptable CLs with 
${{\rm CL}} \ge 0.1\%$, and that the affine linear fits with $\Teff = 
\Tf + m \langle \uT\rangle^2$ also reflect the slopes of the 
single-particle spectra $\Teff$ with a CL~$\geq 0.1$\%.

In addition to the systematic investigations detailed in this 
manuscript, several additional consistency checks were performed, such 
as including fit-range stability investigations and using three 
different methods of propagation of statistical, and systematic 
uncertainties. These were performed because the definition and the 
utilization of $\lambda_{\rm max}$ in this experimental manuscript, 
which has direct access to results within various experimental cuts, 
differs from the utilization of $\lambda_{\rm max}$ in 
Refs.~\cite{Csorgo:2009pa,Vertesi:2009wf}, which reported re-analysis of 
already published data. In those works, only the published systematic 
uncertainties were available to be propagated to the final results, 
while in the present paper every quantity up to the final results has 
been evaluated within each experimental cut.

The method of propagation of statistical and systematic uncertainties 
was cross checked though comparison with the less direct methods of 
Refs.~\cite{Csorgo:2009pa,Vertesi:2009wf} as well as with the PHENIX 
method of Ref.~\cite{PHENIX:2008ove}. These methods gave results that 
are consistent with (and typically have smaller uncertainties than) 
those presented in the body of this manuscript. Hence the central 
values, with statistical and systematic uncertainties, presented in the 
manuscript are obtained with the most conservative of the four different 
methods that were tested.



%
 
\end{document}